\documentclass[apj,twocolumn,twocolappendix,numberedappendix]{openjournal}
\usepackage{newtxtext,newtxmath,enumitem,multirow,ctable,verbatim}
\usepackage[T1]{fontenc}
\usepackage[breaklinks,colorlinks,citecolor=blue,urlcolor=blue]{hyperref}

\newcommand{\Alf}{{Alfv\'en}}

\newcommand{\GIZMO}{{\small GIZMO}}
\newcommand{\gizmourl}{\href{http://www.tapir.caltech.edu/~phopkins/Site/GIZMO.html}{\url{http://www.tapir.caltech.edu/~phopkins/Site/GIZMO.html}}}
\newcommand{\FIREurl}{\href{http://fire.northwestern.edu}{\url{http://fire.northwestern.edu}}}
\newcommand{\paperone}{Paper {\small I}}

\newcommand{\orcidauthor}[3]{\author{\href{http://orcid.org/#1}{#2$^{#3}$}}}

\shorttitle{Radiation \&\ Outflows From ISCO to ROI}
\shortauthors{Hopkins et al.}

\begin{document}

\title{\vspace{-0.8cm}Zooming Out From Magnetically-Dominated Quasar Disks:\\ Radiation and Outflows From ISCO to Parsec Scales\vspace{-1.5cm}}

\orcidauthor{0000-0003-3729-1684}{Philip F. Hopkins}{1,*}
\orcidauthor{0009-0002-8417-4480}{Jaeden Bardati}{1}
\orcidauthor{0000-0002-2858-9481}{Hyerin Cho}{1,3}
\orcidauthor{0000-0003-1598-0083}{Kung-Yi Su}{2}

\affiliation{$^{1}$TAPIR, Mailcode 350-17, California Institute of Technology, Pasadena, CA 91125, USA}
\affiliation{$^{2}$CIERA and Department of Physics and Astronomy, Northwestern University, 1800 Sherman Ave, Evanston, IL 60201, USA}
\affiliation{$^{3}$Canadian Institute for Theoretical Astrophysics, 60 St. George St, Toronto, ON M5S 3H8, Canada}

\thanks{$^*$E-mail: \href{mailto:phopkins@caltech.edu}{phopkins@caltech.edu}},

\begin{abstract}
Simulations have begun to zoom in from scales of galaxies to event horizon ($R_{g}$; au) around supermassive black holes (BHs), to follow quasar evolution self-consistently. These predict novel phenomena such as hyper-magnetized accretion disks, and can uniquely address how feedback generated on small scales reaches the ISM. But previous simulations were limited in evolution by numerical timesteps at small scales and limited physics at large scales, so feedback+radiation could not reach from small-to-large scales. We extend these with time-dilation methods to evolve orders of magnitude longer to $\sim 10^{9.5} G M_{\rm BH}/c^{3} \sim 10^{4}\,$yr, long enough for AGN radiation/outflows to reach the BH radius of influence (ROI; few pc). We study Eddington-scaled accretion rates $0.001\lesssim\dot{m}\lesssim30$, and quantify escaping radiation and outflows. Radiative efficiencies $\epsilon_{r}$ decrease $\propto\dot{m}$ at low-$\dot{m}$ and $\propto\dot{m}^{-1}$ at $\dot{m}\gg1$, peaking at $\sim 0.1$ in-between. At high-$\dot{m}$, inflow rates vary weakly from ROI to horizon in strongly-magnetized disks. The radiation is primarily thermal with $T_{\rm rad,\,eff}\sim 10^{4}-10^{5}\,$K with infrared re-emission from the 'torus,' with illuminated outer disks. A broad-angle radiation-pressure driven outflow is launched with mass-loading at the ROI a few times $\dot{M}_{\rm BH}$ and momentum-loading $\sim L/c$ when $\dot{m}<1$ increasing up to a saturation level $\sim 10\,L/c$ when $\dot{m} \gg 1$ (where $\epsilon_{r} \ll 1$). Velocities span $\sim 10^{2}-10^{5}\,{\rm km\,s^{-1}}$ akin to NLR/BLR/BAL/UFO outflows.  At low $\dot{m}\sim0.001$ this has transitioned to a hot, optically-thin, radiatively-inefficient inner accretion flow with decreasing inflow ($\dot{M}\propto r^{1/2}$). Radiation pressure no longer launches outflows but slow convective flows remain.
\end{abstract}


\maketitle

\section{Introduction}
\label{sec:intro}

Accretion onto supermassive black holes (BHs) powers quasars and other active galactic nuclei \citep[AGN][]{schmidt:1963.qso.redshift,salpeter64}, building up most of their present day mass \citep{soltan82} with luminosities and accretion rates ranging from $L_{\rm bol} \gtrsim 10^{46}\,{\rm erg\,s^{-1}}$ and $\dot{M}_{\rm BH} \gtrsim 1\,{\rm M_{\odot}\,yr^{-1}}$  in typical quasars at redshifts $z\gtrsim 2$ to orders of magnitude smaller in low-luminosity AGN \citep{shen:bolometric.qlf.update}. 
The gas which fuels BH accretion comes from the interstellar medium (ISM) of galaxies, beyond the BH radius of influence (BHROI; scales $R_{\rm ROI} \sim G\,M_{\rm BH}/\sigma_{\ast}^{2}\sim $\,a few pc, in terms of the central galaxy potential $\sigma_{\ast}^{2}$ and BH mass $M_{\rm BH}$), and this makes the structure and basic qualitative properties of the accretion flow sensitive to the ``outer boundary conditions'' of the ISM \citep{pringle:accretion.review,frank:2002.accretion.book,abramowicz:accretion.theory.review,jafari:2019.mhd.ppd.review}, but to be accreted the gas must reach scales closer to the event horizon or Schwarzschild radius ($R_{\rm Schw} \equiv 2\,R_{g} \equiv 2\,G M_{\rm BH}/c^{2} \sim$\,au), where most of the accretion energy is liberated. Strong BH ``feedback'' from this accretion energy generated in the inner accretion disk -- in the form of radiation, outflows, and jets -- is observed escaping back to galactic ISM scales \citep[e.g.][]{laor:warm.absorber,crenshaw:nlr,dunn:agn.fb.from.strong.outflows,sturm:2011.ulirg.herschel.outflows,cafg:2012.egy.cons.bal.winds,faucher-giguere:2012.felobal.model,Zakamska2016,Williams2017}, and is now widely believed to have dramatic effects on both galaxies and BHs \citep{silkrees:msigma,king:msigma.superfb.1,dimatteo:msigma,murray:momentum.winds,hopkins:lifetimes.methods,hopkins:lifetimes.obscuration,torrey:2020.agn.wind.bal.gal.fx.fire}. This can explain the remarkable observed co-evolution and tight correlations between BHs and their host galaxies \citep{magorrian,FM00,Gebhardt00,hopkins:bhfp.obs,aller:mbh.esph,kormendy:2011.bh.nodisk.corr} and control the star formation and mass growth of massive galaxies \citep{croton:sam,hopkins:qso.all,hopkins:groups.ell,wellons:2022.smbh.growth,mercedes.feliz:2023.agn.feedback.positive.negative,cochrane:2023.agn.winds.galaxy.size.effects}. But many of the properties of accretion flows, and how different AGN feedback channels relate to those flows, remain poorly understood both observationally and theoretically.

A major theoretical challenge in understanding these interactions is the extreme dynamic range of scales. The multi-physics, fully non-linear nature of the problem means that numerical simulations are crucial to inform our intuition and make predictions for observations. But from the scalings above, $R_{\rm ROI}/R_{g} \sim (c/\sigma_{\ast})^{2}$, which implies a ratio of characteristic dynamical/evolution timescales ($t_{\rm dyn}(r) \equiv 1/\Omega \sim \sqrt{r^{3}/G M_{\rm BH}}$) of $t_{\rm dyn}(R_{\rm ROI})/t_{\rm dyn}(R_{g}) \sim (c/\sigma_{\ast})^{3}$ or $\gtrsim 10^{9}$ for a typical $\sigma_{\ast} \sim 200\,{\rm km\,s^{-1}}$. And the important physics changes qualitatively over this range as well, especially for high-luminosity/quasar accretion, where at outer radii one must account for star formation and self-gravity with non-equilibrium multi-phase thermochemistry, while at inner radii one must account for strong radiation-matter couplings and magnetic fields. In the last few years there has been dramatic progress on this front. New generations of simulations have (1) expanded the range of physics which can be self-consistently modeled in a single simulation, enabling simulations which encompass all of these limits simultaneously \citep{cho:2023.multiscale.accretion.sims.bondi.inflow.model,cho:2024.multizone.grmhd.sims.bondi.flow.lowaccrate,cho:2025.bridging.scales.jet.host.galaxy.subgrid.model,Cho2026,hopkins:superzoom.overview,hopkins:superzoom.disk,hopkins:multiphase.mag.dom.disks,guo:2024.fluxfrozen.disks.lowmdot.ellipticals}, while also (2) leveraging novel adaptive, super-Lagrangian refinement (and de-refinement) techniques to expand the range of scales surveyed in a single simulation by orders of magnitude, relative to previous efforts \citep{hopkins:zoom.sims,daa:20.hyperrefinement.bh.growth}.

One surprising result from these simulations, and an example of why said simulations are important, has been the discovery of a qualitatively distinct form of accretion disk, so-called ``hyper-magnetized'' or ``flux-frozen'' disks \citep{hopkins:superzoom.disk,hopkins:superzoom.analytic}, wherein the magnetic pressure is much larger than thermal pressure and the field geometry is primarily toroidal. As shown in these works and \citet{hopkins:multiphase.mag.dom.disks}, this predicts orders-of-magnitude different accretion disk properties (e.g.\ masses, thickness/scale-height, star formation rates, accretion \&\ variability timescales) relative to classic thermal-pressure-dominated models like the \citet{shakurasunyaev73} $\alpha$-disk. These emerge as magnetic flux and field lines from the ISM are tidally captured and wrapped around the BH to form the disk, then amplified with the accretion flow \citep{hopkins:superzoom.disk}. Idealized numerical studies \citep{begelman.pringle:2007.acc.disks.strong.toroidal.fields,johansen.levin:2008.high.mdot.magnetized.disks,salveson:2016.sims.mri.dominated.bfield.disks} and examples now seen with a wide variety of distinct numerical methods at exceptionally high resolution \citep{kudoh:2020.strong.b.field.agn.acc.disk.sims.compare,huang:2023.rmhd.sims.rad.similar.mag.pressure,guo:2024.fluxfrozen.disks.lowmdot.ellipticals,kaaz:2024.hamr.forged.fire.zoom.to.grmhd.magnetized.disks,luo:2024.magnetically.dominated.disk.like.our.zoomins.zoomin.on.first.supermassive.star.situation,tsung:2025.magnetic.field.suppresses.fragmentation.agn.disks.strongly.magnetized.idealized.sims,gerling.dunsmore:2025.mhd.agn.disk.sims.strong.b.stabilizes.vs.grav.instab.and.sf.very.strongly.especially.for.toroidal.fields,wang:2025.hypermagnetized.circumbinary.disk.flux.frozen.cavity.to.pc.scales,wang:2025.mad.magnetized.circumbinary.disks.strong.bfields.large.cavities.scale.of.binary} indicate that the formation of these disks is physically robust (not a numerical artifact), though many open questions remain regarding their long-term evolution and how magnetic flux and turbulence are maintained and flow inwards \citep{squire:2024.mri.shearing.box.strongly.magnetized.different.beta.states,guo:2025.idealized.sphere.collapse.sims.hypermagnetized.disks.resolution.dependent.on.resolving.thermal.scale.height.but.limited.physics,tomar.hopkins:2025.lagrangian.vs.eulierian.methods.toroidally.magnetized.disks.guo.test.problem.would.see.collapse.even.low.res.in.mfm.mfv}. Furthermore, observations of broad-line region (BLR), maser, and outer-disk kinematics and magnetic fields appear to strongly favor magnetic pressures much larger than thermal pressures in at least the outer accretion disks (radii $\gg 100\,R_{g}$, with constraints at smaller radii much weaker; \citealt{modjaz:2005.agn.maser.modeling.Bfield.constraints.favor.fluxfrozen.disks,vlemmings:2007.maser.b.limits,gravity:2018.sturm.blr.rotating.thick.disk,gravity:2020.resolved.blr.size.disk.inside.dust.sub,gravity:2021.resolved.blr.disk.hot.dust.coronal.regions,hopkins:2025.maser.evidence.for.mag.disks,takasao:2026.magnetized.black.hole.envelope.model.for.lrds.explains.broad.lines.no.xrays,condo:2026.iras13224.ufo.velocity.luminosity.correlation.favors.magnetic.driving}). Since these disks have received relatively little historical study in simulations, it is especially interesting to ask what their feedback properties should be, and how they connect between small and large scales. 

In this paper, we extend the simulations of \citet{hopkins:multiphase.mag.dom.disks} (henceforth \paperone) to explore these questions. In \paperone, we studied the properties of these disks ``zooming in'' from inter/circum-galactic medium (IGM/CGM) through galaxy and ISM to ROI to near-horizon scales. But while these addressed the dynamic range in space ($R_{\rm ROI}/R_{g}$), they were only run for a time $\sim 10^{4}-10^{5} R_{g}/c \sim 10^{5} G M_{\rm BH}/c^{3}$ ($\sim 5-50$\,days physical time for $M_{\rm BH} \sim 10^{7}\,M_{\odot}$) at maximum refinement, owing to the extremely small timesteps required by the short dynamical and light-crossing times on small scales. Here we extend these simulations using a recently-proposed multi-scale technique \citep{hopkins.most:2025.time.dilation.methods.extreme.timestepping}, to run to $> 10^{9.3} G M_{\rm BH}/c^{3}$ (several thousand years physical time), allowing us to bridge the {\em timescale} gap $t_{\rm dyn}(R_{\rm ROI})/t_{\rm dyn}(R_{g}) \sim (c/\sigma_{\ast})^{3}$ for the first time in simulations of quasar-level systems. 

Our focus here is on the high-luminosity/accretion-rate (quasar) regime, from $\sim 1\%$ to above the Eddington limit. For low-luminosity AGN (LLAGN), with accretion rates many orders of magnitude below Eddington, simulations like those in \citet{cho:2023.multiscale.accretion.sims.bondi.inflow.model,cho:2024.multizone.grmhd.sims.bondi.flow.lowaccrate,cho:2025.bridging.scales.jet.host.galaxy.subgrid.model,Cho2026,guo:2024.fluxfrozen.disks.lowmdot.ellipticals} are leading the way in understanding the BH-ISM coupling. Importantly, the key physics (and therefore validity of different simulations) is quite distinct in these regimes. In the quasar regime, a challenge for decades has been understanding why accretion disks do not fragment and ``shatter'' into stars, so star formation, stellar feedback, and self-gravity are essential; the outer disks are observed to be highly multi-phase, with neutral (atomic \&\ molecular) and dusty phases as well as highly-ionized gas; and the inner regions feature radiation forces comparable to or exceeding gravity, while the disk transitions between optically-thin and thick cooling over the scales of interest, necessitating explicit radiation-magnetohydrodynamics \citep{hopkins:radiation.methods,hopkins:multiphase.mag.dom.disks}. But radio jets appear observationally to be a sub-dominant form of feedback in most luminous quasars \citep{davis:2020.mhd.sim.acc.disk.review}, so we neglect fully self-consistent general-relativistic magnetohydrodynamic (GRMHD) treatments of jets in order to capture the other physics above. For LLAGN, star formation/self-gravity, multi-phase disk structure, and strong radiation-matter couplings are much less important, but jets are much more important, making simulations like \citet{cho:2023.multiscale.accretion.sims.bondi.inflow.model,cho:2025.bridging.scales.jet.host.galaxy.subgrid.model,guo:2024.fluxfrozen.disks.lowmdot.ellipticals,Su2026} that self-consistently join to full GRMHD at near-horizon scales essential to understand LLAGN.

In \S~\ref{sec:sims} we summarize the simulations, methods, and numerical time-dilation scheme. \S~\ref{sec:overview} presents the basic accretion disk properties and their quasi-steady-state behavior. \S~\ref{sec:outflow} examines the outflows and jets across accretion rates, and \S~\ref{sec:rad} the outgoing radiation, including efficiencies, emergent spectra, and illumination of the outer disk. \S~\ref{sec:subgrid} proposes sub-grid prescriptions for feedback from high and low accretion states, and \S~\ref{sec:discussion} discusses caveats and implications. We summarize in \S~\ref{sec:conclusions}.

\begin{figure*}
	\centering\includegraphics[width=0.97\textwidth]{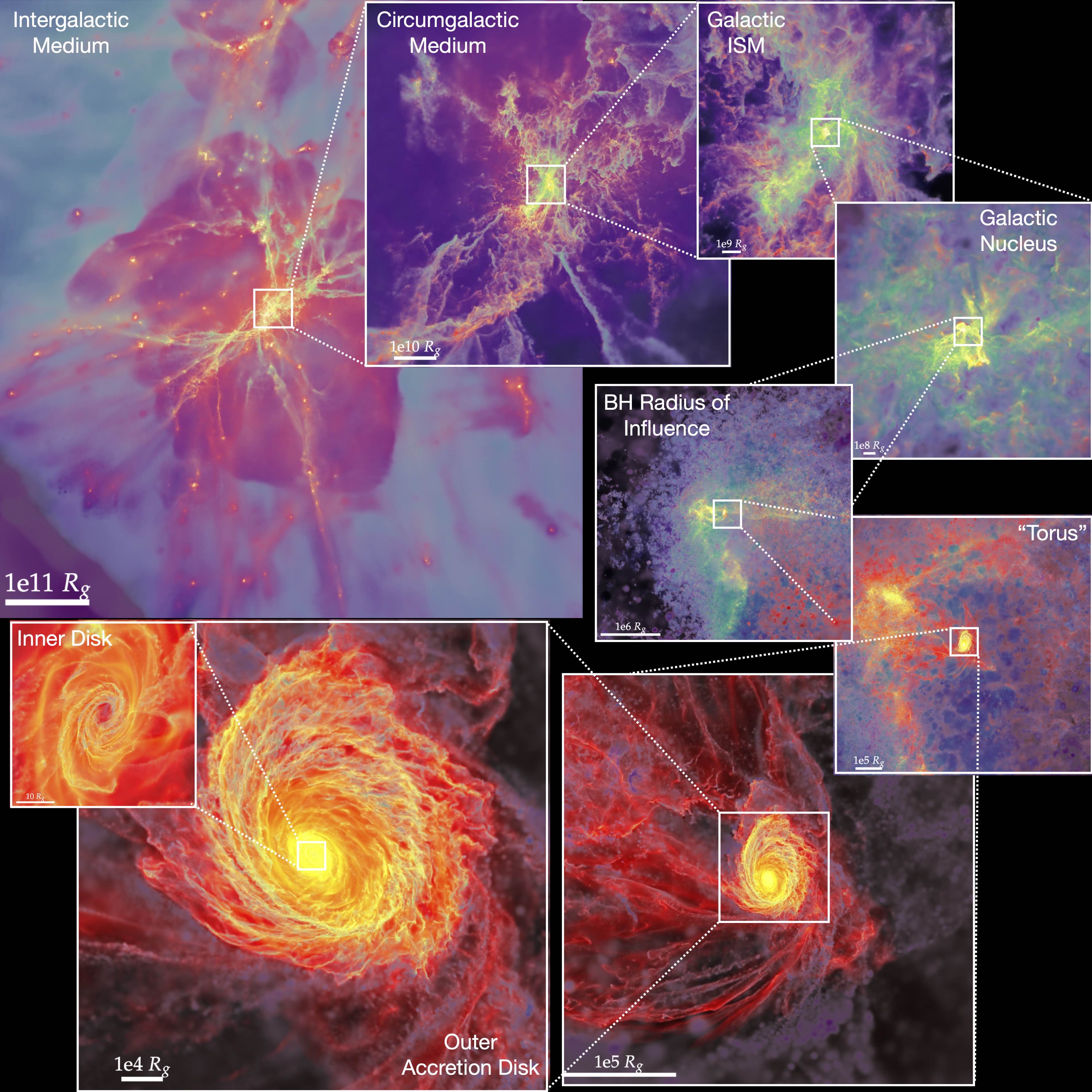} 
	\caption{Images of the ``initial conditions'' (\S~\ref{sec:sims:overview}) reproduced from \paperone. Bar in units of $R_{g} \equiv 2 G M_{\rm BH}/c^{2}$ ($M_{\rm BH}=1.3 \times 10^{7}\,{\rm M_{\odot}}$), intensity scales with log of surface density ($\sim 10^{4}$ range in each) with colormaps showing the dominant pressure (radiation+thermal in {\em purple-magenta}; turbulent/kinetic in {\em blue-green}; magnetic in {\em red-yellow}). The \paperone\ simulation begins from a cosmological box with trace ($\sim 10^{-15}\,$G) magnetic fields at redshift $z\approx150$, and zooms into a massive, multiply-merger starburst galaxy at $z\sim4.5$, where a massive ($\sim10^{8}\,{\rm M}_{\odot}$) GMC complex is disrupted around the BHROI ($\sim 5\,$pc) and forms an accretion disk at sub-pc scales, which \paperone\ zoomed-in on following inflows to $\sim 3\,R_{g}$.\label{fig:zoomies}}
\end{figure*}

\begin{figure}
	\centering	\includegraphics[width=0.98\columnwidth]{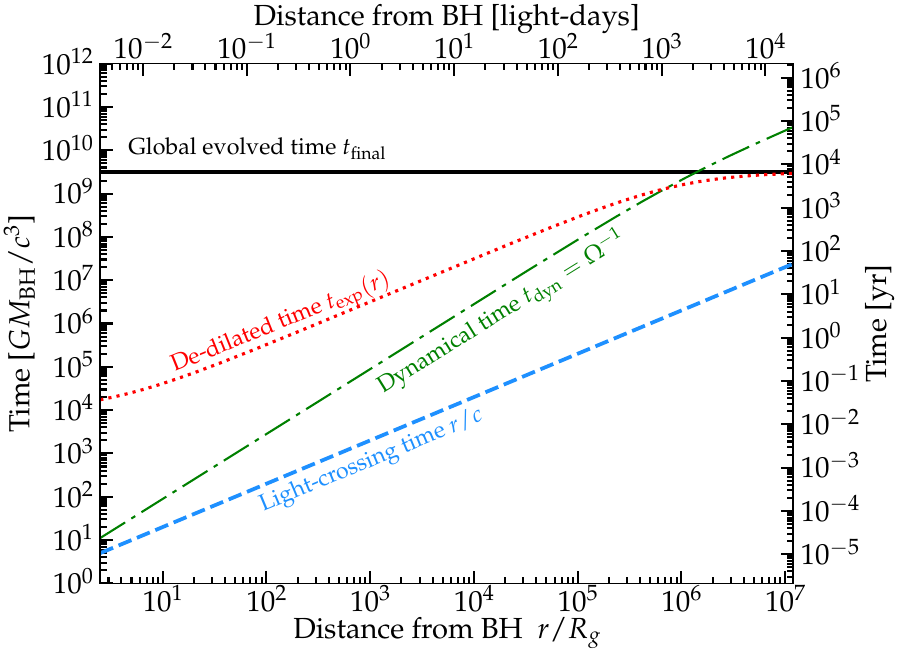}
	\caption{Characteristic timescales vs.\ BH-centric radius in these simulations. We compare dynamical time $t_{\rm dyn}\equiv \Omega^{-1} \equiv r/V_{c} \equiv \sqrt{r^{3}/GM(<r)}$, light-crossing time, the global evolved simulation time $10^{9.5}\,G {\rm M}_{\rm BH}/c^{3}$, and the ``de-dilated'' time $t_{\rm exp}(r) =a(r)\, t_{\rm final}$ (the effective physical time each radius experiences if we remove the dilation factor; \S~\ref{sec:sims:dilation}). Most radii from ISCO to ROI ($\sim 4000\,{\rm ld}$) are evolved multiple light-crossing and dynamical times. We focus on this range of radii in what follows given our finite time-evolution.\label{fig:profiles.timescales}}
\end{figure}

\section{Simulations \&\ Methods}
\label{sec:sims}

The simulations we study are a continuation of those in \paperone, themselves an extension of those studied in \citet{hopkins:superzoom.overview,hopkins:superzoom.disk,hopkins:superzoom.imf}. All salient details of the physics, numerical methods, and initial conditions are identical to those therein, excepting the time-dilation scheme detailed below, so we only briefly review the former.

\subsection{Overview, Physics, \&\ Initial Conditions}
\label{sec:sims:overview}

The simulations (Fig.~\ref{fig:zoomies}) begin with a $\sim (100\,{\rm cMpc})^{3}$ periodic box at redshift $z\sim 150$, with trace primordial magnetic fields ($\sim 10^{-15}\,$G), and evolve as a cosmological galaxy formation simulation with the full combined Feedback In Realistic Environments\footnote{\FIREurl} (FIRE) and STARFORGE physics \citep{hopkins:fire2.methods,hopkins:fire3.methods,grudic:starforge.methods,guszejnov:2020.starforge.jets}, in the code GIZMO\footnote{A public version of the code is available at \gizmourl} \citep{hopkins:gizmo}. At $z\sim 4.5$, when the galaxy is a massive, clumpy, starburst (with star formation rate $\sim 200\,{\rm M}_{\odot}\,{\rm yr^{-1}}$), a massive molecular cloud complex is perturbed to encounter the supermassive black hole (mass $M_{\rm BH} \sim 1.3\times10^{7}\,{\rm M_{\odot}}$), and the simulation is hyper-refined to follow the gas inflows in the central $\lesssim 10-100\,$pc, reaching $< 0.005\,{\rm M}_{\odot}$ resolution interior to the ROI ($\sim 5\,$pc). Tidally stripped material from the disrupted cloud circularizes at $\sim 0.1-1\,$pc, forming an accretion disk \citep{hopkins:superzoom.disk,hopkins:superzoom.analytic} which is followed in \paperone\ down to the ISCO. The refinement scheme resolves the disk with roughly constant ${\rm d} N_{\rm cells}/{\rm d}\ln{R} \sim 10^{7}$ cells per log-radius (see \paperone, \S~2, Figs.~3 \&\ 16), with mass resolution scaling as $\Delta m \propto r^{2.4}$ interior to $\sim 10^{4} R_{g}$, producing a minimum (median) midplane spatial resolution $\Delta x \approx 10^{-4}\,r$ ($\sim 0.002\,r$). The \paperone\ simulations were gradually refined (running for many dynamical times at each radius while ``zooming in''), then run for $\sim 10^{4}-10^{5} G M_{\rm BH}/c^{3}$ at maximum refinement; the final snapshot of these forms the initial conditions here.

The physics evolved include magnetohydrodynamics \citep[following][]{hopkins:mhd.gizmo,hopkins:cg.mhd.gizmo}, with anisotropic viscosity and conduction \citep{su:2016.weak.mhd.cond.visc.turbdiff.fx,hopkins:gizmo.diffusion}; self-gravity (with adaptive self-consistent softening, scaling with the resolution, and a high-order Hermite integrator for orbital dynamics; \citealt{grudic:2020.tidal.timestep.criterion,grudic:starforge.methods,grudic:2021.accelerating.hydro.with.adaptive.force.updates,hopkins:tidal.softening}); metal enrichment/diffusion and dust destruction/sublimation \citep{ma:2016.disk.structure,gandhi:2022.metallicity.dependent.Ia.rates.statistics.fire,choban:2022.fire.dust.growth.destruction.chemistry}; (proto)star formation and accretion and explicit feedback from stars in the form of protostellar jets, main-sequence stellar mass-loss, radiation, and supernovae \citep{grudic:2022.sf.fullstarforge.imf,guszejnov:2022.starforge.cluster.assembly,guszejnov:environment.feedback.starforge.imf,guszejnov:starforge.environment.multiplicity}; multi-group, fully-NLTE M1 radiation-hydrodynamics (RHD; with fixed FIR, OIR, NUV, EUV, and ionizing plus adaptive-wavelength bands with explicitly evolved $T_{\rm rad}({\bf x},\,t)$ in each cell), coupled to all thermo-chemical processes \citep{hopkins:radiation.methods,hopkins:2019.grudic.photon.momentum.rad.pressure.coupling,grudic:starforge.methods}; non-equilibrium thermo-chemistry, opacities, and radiative cooling including an extensive set of reactions for multi-phase dusty/molecular/atomic/partially-ionized/metal-enriched media designed to encompass the dominant processes at densities $n \sim 10^{-8} - 10^{16}\,{\rm cm^{-3}}$ and temperatures $\sim 1-10^{10}\,$K (\citealt{hopkins:fire3.methods,lenzuni:1991.opacities,semenov:2003.dust.opacities,john:1988.hminus.opacity,glover:2007.low.metallicity.cooling.h2.tables} and \paperone). Note that the RHD treatment includes terms up to and including $\mathcal{O}(v_{\rm gas}^{2}/c^{2})$ in radiation transport and radiation-matter couplings \citep{mihalas:1984oup..book.....M}, but our broad-band treatment means that we do not capture resonant line scattering. Like \paperone\ there is no reduced-speed of light approximation employed here. Gravity is Newtonian, except we approximate the BH with a Paczy{\'{n}}ski-Wiita (PW) potential; lacking a fully general relativistic (GR) treatment, the inner boundary is a simple sink at the ISCO. 

The initial result from this refinement is a system accreting super-critically with $\dot{m} \sim 20$ ($\dot{M}_{\rm BH} \sim 5-10\,{\rm M_{\odot}\,yr^{-1}}$) at the ISCO.\footnote{We define the dimensionless accretion rate $\dot{m} \equiv \dot{M}_{\rm BH} / \dot{M}_{\rm crit} \equiv 0.1\,\dot{M}_{\rm BH} c^{2} / L_{\rm Edd}$, i.e.\ relative to Eddington for a reference radiative efficiency $0.1$, so for this BH mass $\dot{M}_{\rm crit} \approx 0.3\,{\rm M_{\odot}\,yr^{-1}}$.} To compare accretion rates in a controlled manner, \paperone\ re-started the refinement from the BHROI after re-scaling the mass of each gas cell by a factor $f_{\dot{M}}$ (keeping BH and star/dark matter masses fixed) and re-scaling the initial magnetic and thermal pressures accordingly (to maintain fixed $H/R$ and the captured-gas virial parameter and plasma $\beta$). Three simulations were run, $f_{\dot{M}} = (1,\,0.03,\,0.001)$, corresponding in \paperone\ to accretion rates $\dot{m} \sim (15-30,\,0.3-0.4,\,0.02-0.04)$ or $\dot{M}_{\rm BH} \sim (5,\,0.1,\,0.007)\,{\rm M_{\odot}\,yr^{-1}}$. The intermediate case, being most representative of ``typical'' quasar properties, forms our fiducial case-study here; we discuss the other two primarily where they differ qualitatively.

\subsection{Extending the Simulations In Time}
\label{sec:sims:dilation}

We re-start from the final times in \paperone, activating the ``time-dilation'' scheme from \citet{hopkins.most:2025.time.dilation.methods.extreme.timestepping} (hereafter HM25). As described therein, this is a very simple modification of the numerical system: for all the evolved variables ${\bf U}_{i}$ in each cell $i$, their evolution equations in GIZMO are modified to $D {\bf U}_{i} / D t \rightarrow a_{i}^{-1} D {\bf U}_{i} / D t$ with $a_{i}({\bf x} = {\bf x}_{i},\,t = t_{i}^{(n)})$ the dimensionless dilation/scale/stretch factor defined below. As shown in HM25 this is a straightforward generalization of methods already widely used for multi-scale problems such as the reduced-speed-of-light method for RHD \citep{gnedin:2016.reduced.c.scale.separation.background.vs.time,rosdahl:2018.sphinx.sims.binaries.reduced.c.variable,ji:2021.mhd.pic.rsol} or ``slowdown'' methods for collisional N-body simulations \citep{mikkola.aarseth:1996.slowdown.for.close.binaries.dynamics}. Effectively, the dynamics in regions with $a_{i} \ll 1$ are ``slowed down'' by a factor $a_{i}$, enabling them to be advanced along the global timeline by larger timesteps $\Delta t \rightarrow a_{i}^{-1} \Delta t$, which is valid so long as the system is in some kind of statistical quasi-steady-state on scales where $a_{i} \ll 1$. 

We have tested various schemes for $a_{i}$ (with ``no dilation,'' e.g.\ the \paperone\ simulations, corresponding to $a_{i} = 1$), but in our ``default'' simulations we adopt (for most of their run-time) 
\begin{align}
\label{eqn:dilation.adopted} a_{i}^{-1} \approx 1 + \left[ a_{\rm min} + (r / r_{\rm out})^{\alpha_{a}} \right]^{-1}
\end{align}
where $r_{\rm out} \equiv a_{\rm min}^{-1} r_{\rm ISCO} = a_{\rm min}^{-1} 6\,G M_{\rm BH} / c^{2}$, and the default $a_{\rm min} \approx 2\times10^{-6}$, $\alpha_{a} \approx 1$. This simply interpolates between $a_{i} \rightarrow a_{\rm min}$ at $r\lesssim r_{\rm ISCO}$ and $a_{i} \rightarrow 1$ at $r \gtrsim r_{\rm out}$, with $a_{i} \approx (r/r_{\rm out})^{\alpha_{a}}$ in-between. With $\alpha_{a} \sim 1$, gas closer to the BH still evolves ``faster'' (with smaller timescales and timesteps) than gas further out, but the imbalance is reduced: the dynamic range of numerical timesteps from \S~\ref{sec:intro} is transformed from $\sim (c/\sigma_{\ast})^{3} \rightarrow (c/\sigma_{\ast})$, enabling us to run $10^{4}-10^{5}$ times longer for a CPU cost similar to the original \paperone\ simulations. The resulting effective (``de-dilated'') evolution time at each radius is compared to the local dynamical and light-crossing times in Fig.~\ref{fig:profiles.timescales}. 

This choice is motivated by (1) maintaining the conditions for numerical stability and accuracy defined in HM25; (2) scalings which produced good behavior on the idealized and multi-physics test problems therein; and (3) restricting the dilation to the Keplerian region ($r_{\rm out} \sim R_{\rm ROI}$), both because that is the region studied here and because reference scalings like the dynamical time are simple power laws about a single well-defined point (the BH).

Note that for outflow or radiation moving radially outwards from some launching radius $r_{i}$ to a final $r_{f}$ at constant speed $v_{\rm out}$, the global time required to reach $r_{f}$ with dilation is $t_{f}= \int dr/(a\,v_{\rm out})$, so instead of $t_{f} = (r_{f}-r_{i})/v_{\rm out}$, for our default $a \approx (r/r_{\rm out})$ and $r_{\rm out}\sim r_{f}$ this becomes $t_{f} \approx r_{\rm f}/v_{\rm out}\,\ln{(r_{f}/r_{i})}$. So material with $v_{\rm out} \gtrsim 1000\,{\rm km\,s^{-1}}$ (most of the outflows of interest, and radiation so long as the trapping optical depth is $\tau \lesssim 10^{4}$) can propagate from ISCO to beyond the ROI in the time we simulate. Slower/more trapped outflows would require longer evolution to cover the entire domain, but can be at least partially represented by the alternative simulations described below.

\subsection{Tradeoffs of this Approach \&\ Validation}
\label{sec:sims:tradeoffs}

As emphasized in HM25, ``nothing is free'': the major limitation of a dilation factor like Eq.~\ref{eqn:dilation.adopted} is that it is only strictly valid when the system is in local (statistical) steady-state, and the further from equilibrium, the less accurate the evolution relative to the ``un-dilated'' case. Because we explicitly slow down the small-scale evolution, we also sacrifice meaningful predictions for time-variability on scales $\ll a_{i}^{-1} \Omega^{-1}$ at each radius. Other bridging scales methods also share this limitation in capturing time variability (e.g., see the extended discussion in \citealt{cho:2024.multizone.grmhd.sims.bondi.flow.lowaccrate}).

To understand how this may influence our results, we experimented with other choices for $a_{i}$, varying $a_{\rm min}$ from $10^{-6}-1$ and $\alpha_{a} = (1/6,\,1/3,\,1/2,\,1,\,5/4)$ in shorter experiments (several shown in HM25). We also ran a parallel set of simulations (for each $f_{\dot{M}}$) with less-aggressive dilation $(a_{\rm min},\,\alpha_{a})=(10^{-4},\,1)$ to $\gtrsim 10^{7} G M_{\rm BH}/c^{3}$ to compare convergence with respect to $a_{i}$. And for the $f_{\dot{M}}=1$ run we periodically ``de-dilated'' the simulation following HM25, restarting from a snapshot and either turning off dilation ($a_{i} \rightarrow 1$) or ``ramping'' it down (decreasing $r_{\rm out}$ as ${\rm d} r_{\rm out} /{\rm d} t = -0.1\,r_{\rm out} \Omega(r_{\rm out})$), to check that no unphysical behaviors were appearing and that the quasi-steady behavior at small radii was maintained for some number of local dynamical times without dilation. The detailed, quantitative validation of the method itself -- including direct comparisons of dilated vs.\ un-dilated evolution on idealized tests and a multi-physics AGN-disk setup essentially identical to the one here -- is presented in HM25, so here we focus on the physical results, noting where the dilation could plausibly affect a given conclusion.

We also ran a more traditional ``sub-grid model'' or ``zoom-out'' version of the $f_{\dot{M}}=1$ case, reversing the initial zoom-in: we restart with the sink radius increased by $10\times$ and run for $\sim 10-100$ orbital times at the new sink radius, with the inner boundary still a sink but also injecting mass at a constant angle-dependent rate $\dot{M}_{\rm out}(R,\,\theta)$ (following the sub-grid jet implementation in \citealt{su:2021.agn.jet.params.vs.quenching,grudic:starforge.methods}, with opening angle $\theta_{\rm out}$ and velocity $\langle v_{\rm out}(r) \rangle$) and radiation $L_{\rm out}(R,\,\theta) = L_{\rm out}(r)\,|\cos{\theta}|^{\alpha_{L}}$, all measured and fit from the last snapshot before ``de-refinement''; we then iterate this twice more to reach larger radii. This sacrifices the ability to evolve small and large scales simultaneously, and resembles common sub-grid AGN feedback models, but allows us to check for numerical artifacts of the dilation method. Both the weaker-dilation and zoom-out runs recover the same qualitative $\dot{M}(r)$ and $\rho(r)$ profiles as our default runs (a nearly radius-independent $\dot{M}_{\rm in}$ and steeply-declining $\rho$), although since these are realized at different, earlier times and with different inner boundaries, the agreement should be read at the qualitative time-averaged, not snapshot-by-snapshot, level.

\subsection{Definitions \&\ Reference Values}
\label{sec:definitions}

The BH mass here, $M_{\rm BH} \approx 1.3 \times 10^{7}\,{\rm M_{\odot}}$, gives $L_{\rm Edd} \sim 1.5 \times 10^{45}\,{\rm erg\,s^{-1}}$ and $\dot{M}_{\rm crit} \equiv L_{\rm Edd}/0.1 c^{2} \sim 0.26\,{\rm M_{\odot}\,yr^{-1}}$. We define the gravitational radius $R_{g} \equiv R_{\rm Schw} =  2\,G M_{\rm BH} / c^{2} \sim 0.26\,{\rm au}$ and timescale $t_{g} \equiv G M_{\rm BH} / c^{3} = 64\,{\rm s} \approx 1\,$minute; dimensionless accretion rate $\dot{m} \equiv 0.1\,\dot{M}_{\rm BH} c^{2} / L_{\rm Edd}$; plasma $\beta \equiv P_{\rm therm,\,gas} / P_{\rm mag} = (n k_{B} T) / (B^{2} / 8\pi)$ and ``total'' plasma $\beta_{\rm tot} \equiv (P_{\rm therm,\,gas} + P_{\rm rad}) / (B^{2} / 8 \pi)$ where we take $P_{\rm rad} \equiv e_{\rm rad} / 3$ (regardless of whether the photons are tightly-coupled). We use $(r,\,\phi,\,\theta)$ and $(R,\phi,\,z)$ to refer to spherical/cylindrical coordinates, respectively ($\theta$ the polar angle). Quantities like $\dot{M}_{\rm in}$, $\dot{M}_{\rm out}$ refer to inflow/outflow components, calculated by summation/averaging over all cells within some radial shell with $v_{r} < 0$ or $v_{r} > 0$, respectively. Unless otherwise specified $\dot{M}_{\rm BH}$ refers to the inflow/accretion rate onto the BH at our inner (ISCO) boundary. The circular velocity and angular frequency $V_{c}(r) \equiv \sqrt{G M_{\rm enc}(<r) / r}$, $\Omega(R) \equiv V_{c}(R)/R$ include all enclosed mass so begin to differ from their Keplerian ($M_{\rm enc} \rightarrow M_{\rm BH}$) values as we approach the ROI, $R_{\rm ROI} \sim G M_{\rm BH} / \sigma_{\ast}^{2} \sim 5\,{\rm pc} \sim 4 \times 10^{6} R_{g}$ ($\sigma_{\ast} \sim 100\,{\rm km\,s^{-1}}$ here). Per \citet{hopkins:superzoom.overview} (Fig.~7 therein), interior to $\lesssim 0.5\,{\rm pc}$ gas dominates the (non-BH) mass budget, while exterior to $\sim 0.5\,$pc (out to $\sim$\,kpc) stars dominate, with $\Sigma_{\ast} \sim 10^{6}\,{\rm M_{\odot}\,pc^{-2}}$ at $\sim 0.1-10\,$pc, mostly in a young nuclear star cluster forming the ``core'' of the proto-bulge (stellar mass $\sim 6\times 10^{7}\,{\rm M_{\odot}}$ interior to $\lesssim 10\,$pc). The disk circularization radius is $\sim 0.1\,$pc (for material stripped from the disrupted GMC complex, of total gas mass $\sim 10^{8} {\rm M_{\odot}}$), and the radii we study span our inner boundary (the ISCO) to just outside the ROI, $\sim 3 \rightarrow 10^{7}\,R_{g}, \sim R_{\rm ISCO} \rightarrow 3\,R_{\rm ROI}, \sim 10^{13} \rightarrow 10^{19.6}\,{\rm cm}$.

\begin{table*}
\centering
\caption{Summary of the three fiducial simulations (all at $M_{\rm BH}\approx1.3\times10^{7}\,M_{\odot}$). Values are time-averaged over the dilation-extended run; the inner disk refers to $\lesssim10^{3}\,R_{g}$. Quantities carry $\sim$\,factor-of-few time variability (\S~\ref{sec:rad:eff}) and the low-$\dot{m}$ column uses the $w$Jet run.}
\label{tbl:runs}
\begin{tabular}{lccc}
\hline\hline
Quantity & $f_{\dot{M}}=1$ & $f_{\dot{M}}=0.03$ & $f_{\dot{M}}=0.001$ \\
\hline
Accretion rate $\dot{m}\equiv0.1\,\dot{M}_{\rm BH}c^{2}/L_{\rm Edd}$ & $\sim20$ (super-Edd.) & $\sim0.3$ & $\sim10^{-3}$ \\
$\dot{M}_{\rm BH}$ [$M_{\odot}\,{\rm yr^{-1}}$] & $\sim5$ & $\sim0.1$ & $\sim3\times10^{-4}$ \\
Luminosity $L/L_{\rm Edd}$ & $\sim2$ & $\sim0.3$ & $\sim10^{-4}$ \\
Radiative efficiency $\epsilon_{r}\equiv L/\dot{M}_{\rm BH}c^{2}$ & $\sim0.004$ & $\sim0.09$ & $\sim0.009$ \\
Disk plasma $\beta\equiv P_{\rm th}/P_{\rm mag}$ & $\sim10^{-3}$ & $\sim10^{-3}$ & $\sim1$ \\
Disk aspect ratio $H/R$ & $\sim0.1-0.4$ & $\sim0.05-0.2$ & $\sim0.3-0.5$ \\
$\dot{M}_{\rm in}(r)$ scaling (ROI$\rightarrow$ISCO) & $\propto R^{0.07-0.12}$ & $\propto R^{0.1-0.15}$ & $\propto R^{1/2}$ \\
Dominant inner-disk emission & big blue bump ($10^{4-5}\,$K) & big blue bump & non-thermal X-ray ($\gtrsim10\,$keV) \\
Dominant outflow channel & rad.-pressure, $\dot{P}\sim10\,L/c$ & rad.-pressure, $\dot{P}\sim L/c$ & turbulent/convective + jet \\
Outflow speed [${\rm km\,s^{-1}}$] & $\sim10^{4}-10^{5}$ & $\sim5\times10^{3}-3\times10^{4}$ & $\sim v_{\rm esc}(R_{\rm ROI})$ \\
Accretion state & magnetic disk & magnetic disk & hot MAD/ADAF + cold outer disk \\
\hline
\end{tabular}
\end{table*}

\begin{figure}
	\centering	\includegraphics[width=0.98\columnwidth]{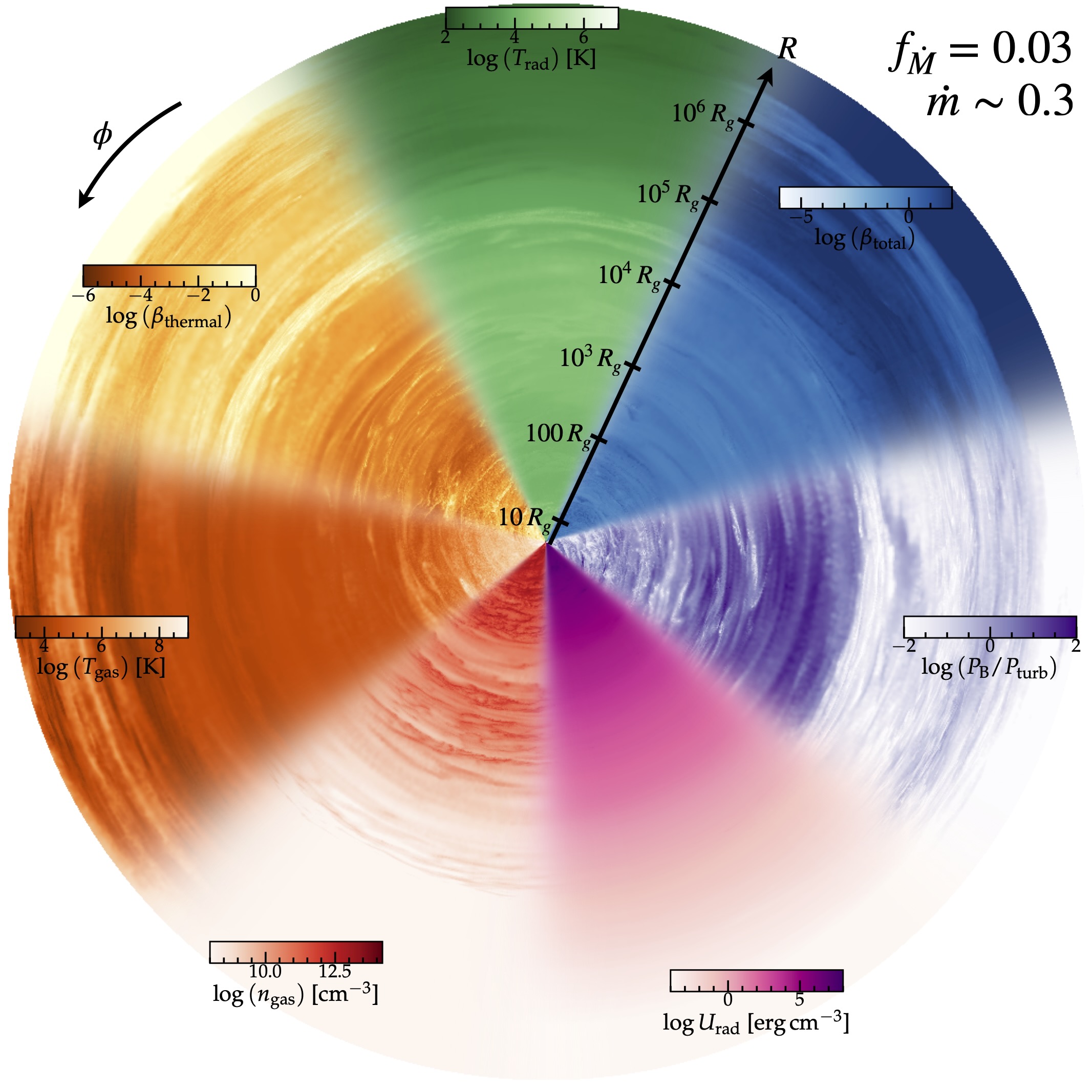}\\\includegraphics[width=0.95\columnwidth]{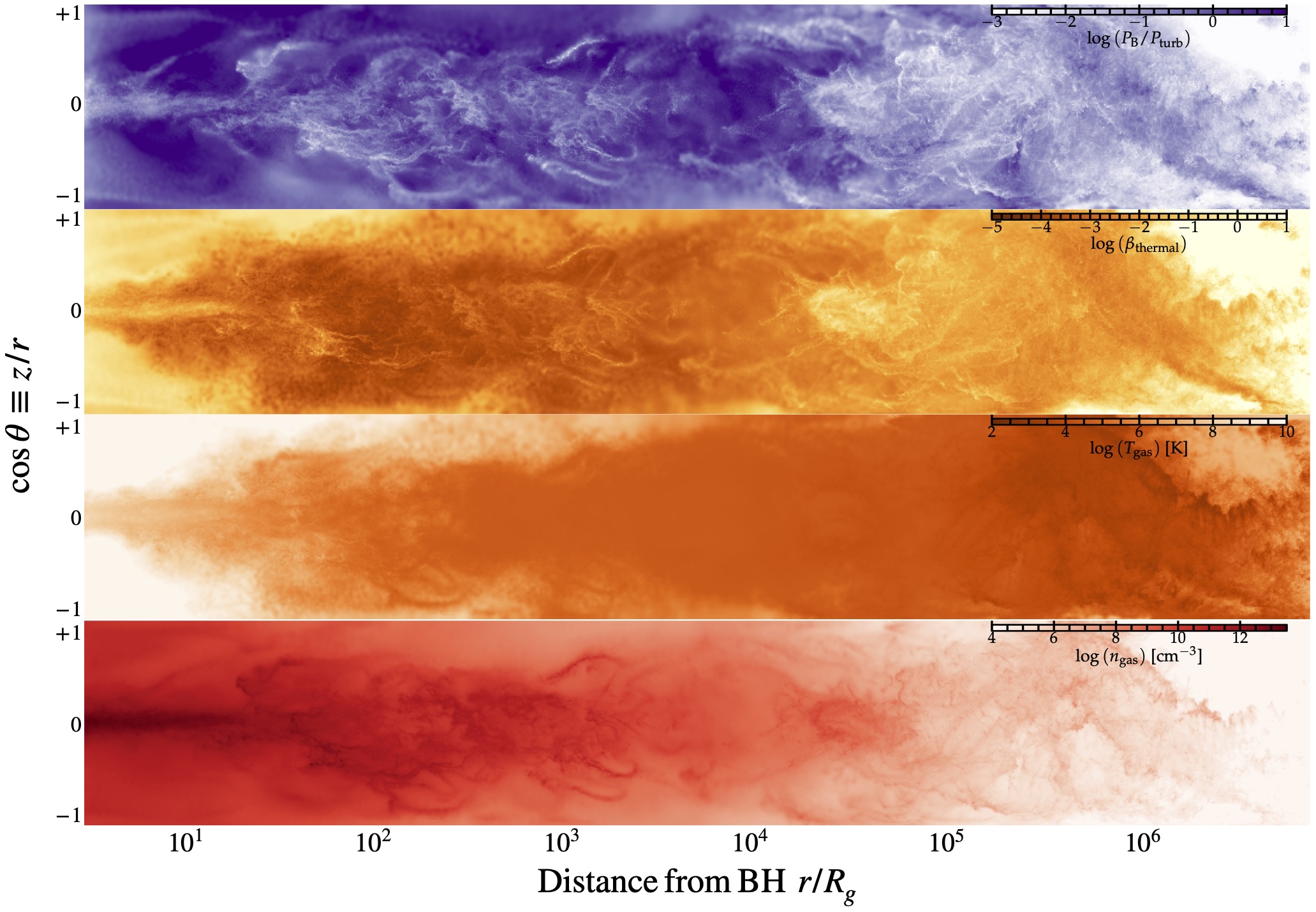}
	\caption{Face on ``pinwheel'' view (as \paperone) of the fiducial $f_{\dot{M}}=0.03$ disk, with BH-centric radius increasing outward from $\sim 3\,R_{g}$ (center) to $\sim 10^{7}\,R_{g}$ (edge) and azimuth $\phi$ around. The angular sectors render different midplane quantities of the final snapshot, each with its own colorbar: radiation temperature $\log T_{\rm rad}$, total plasma $\log\beta_{\rm tot}$, magnetic-to-turbulent pressure $\log(P_{B}/P_{\rm turb})$, radiation energy density $\log u_{\rm rad}$, gas density $\log n_{\rm gas}$, and thermal plasma $\log\beta_{\rm thermal}$. Below we show edge-on $\cos{\theta}-r$ projections of a subset of quantities where the disk vertical structure is evident.\label{fig:maps.pinwheel}}
\end{figure}

\begin{figure*}
	\centering\includegraphics[width=0.32\textwidth]{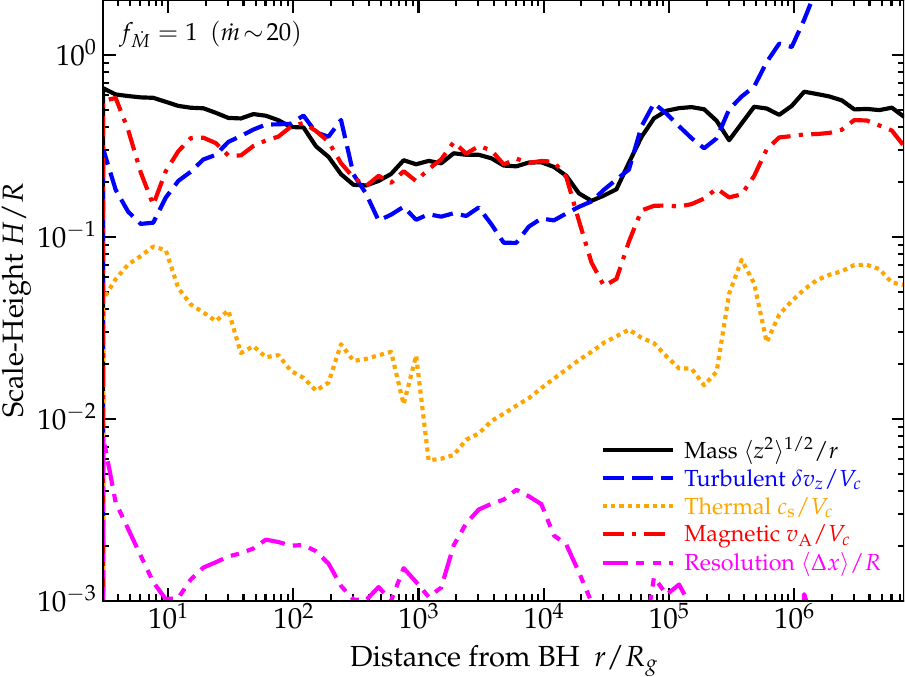}\hspace{2pt}\includegraphics[width=0.32\textwidth]{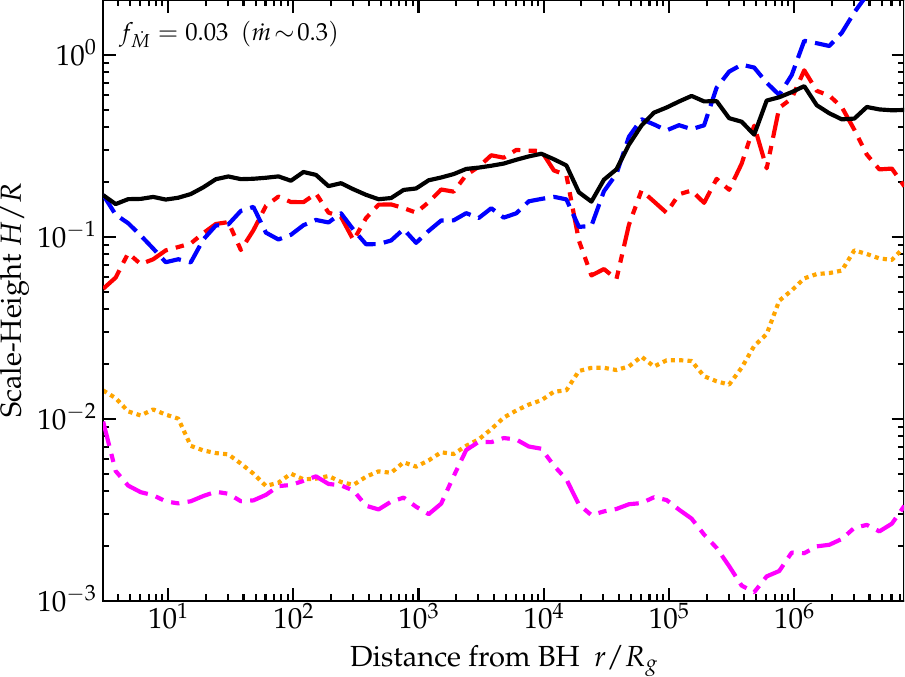}\hspace{2pt}\includegraphics[width=0.32\textwidth]{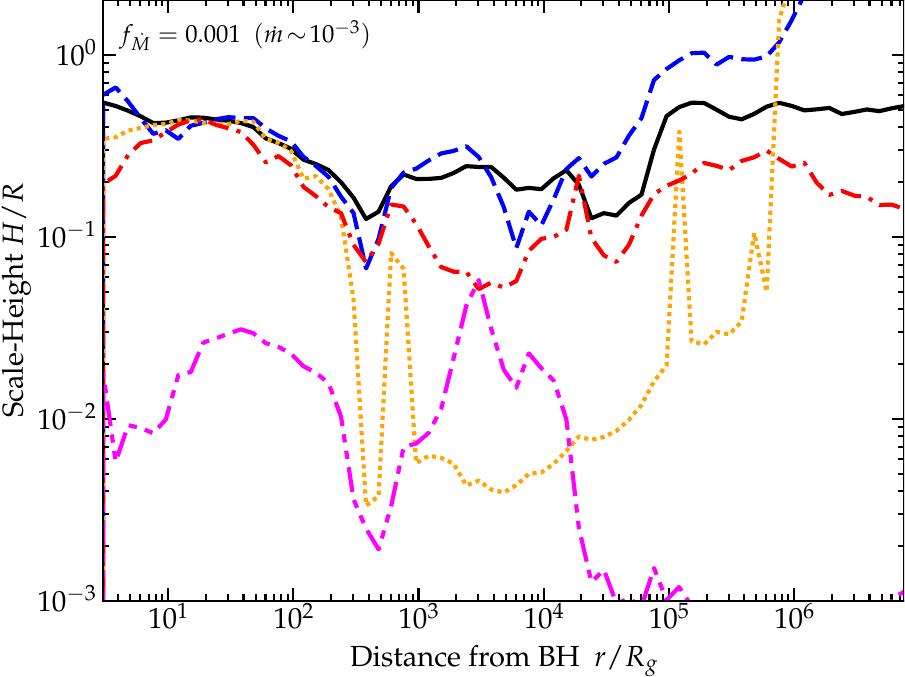}
	\caption{Breakdown of disk vertical support vs.\ radius for the three accretion-rate cases ($f_{\dot{M}}=1,\,0.03,\,0.001$, {\em left to right}). In each panel: RMS geometric aspect ratio $\langle z^{2} \rangle^{1/2}/r$, turbulent vertical velocity $\delta v_{z}/V_{c}$, \Alf\ speed $v_{A}/V_{c}$, thermal sound speed $c_{s}/V_{c}$, and minimum spatial resolution $\langle \Delta x \rangle/R$. High $\dot{m} \gg 0.01$ exhibits slim-to-thick disks supported by magnetic fields+trans-\Alf{ic} turbulence; at low $\dot{m} \ll 0.01$ there is an outer low-$\beta$ region but the disk is very thick with $\beta\sim 1$ and near-virial thermal pressure. All simulations attempt to resolve the thermal height over most of the disk.\label{fig:height.breakdown}}
\end{figure*}

\section{Accretion Disk Properties and Scalings}
\label{sec:overview}

Figs.~\ref{fig:maps.pinwheel}--\ref{fig:height.breakdown} illustrate the disk structure (for the fiducial $f_{\dot{M}}=0.03$ case, and the vertical support in all three runs), and Table~\ref{tbl:runs} summarizes the key time-averaged properties of the three runs. Much of this disk structure is qualitatively unchanged from \paperone, so we focus on what differs at late times.

\subsection{Do The Disks Reach (Quasi) Steady-State?}
\label{sec:steady.state}

Within the caveats of \S~\ref{sec:sims:tradeoffs}, the disks do appear to approach a statistical quasi-steady-state. This is evident in the time series of $\dot{M}$ (Fig.~\ref{fig:mdot}; likewise for $L$, \S~\ref{sec:rad:geom.time}): at $r \lesssim 10^{5}\,R_{g}$, the inflow/outflow rates fluctuate about well-defined means rather than secularly drifting. The details of net inflow-vs-outflow, especially where the two are comparable, can change, and there are significant excursions, but many of the accretion properties (with the exceptions discussed below) in the \paperone\ simulations run to $\sim 10^{4-5}\,t_{g}$ are reasonably predictive of those run to $> 10^{9}\,t_{g}$, at least at radii which the radiation and outflows could already reach in the earlier runs.

We also see no dramatic difference between runs with different dilation schemes (\S~\ref{sec:sims:tradeoffs}), again restricting to radii which the outflows and radiation could reach in a given time. The ``weaker'' dilation runs can only be run to $\sim 10^{7}\,t_{g}$, so their outermost radii are effectively ``frozen'' at their initial state, but in the inner regions both the weaker-dilation and no-dilation (\paperone) runs are within the range of behaviors seen over $\sim 10^{9}-10^{10}\,t_{g}$. Likewise, the ``zoom-out'' runs of \S~\ref{sec:sims:tradeoffs} give outflow and disk properties at $\gtrsim 10^{4}\,R_{g}$ consistent with the dilation runs within their time variability -- the same broad-angle bipolar outflows, similar mass-loading, and comparable radiation flux profiles -- giving additional confidence that our results are not dominated by artifacts of the dilation method.

Even with the suppression of variability implied by our method, there is clear strong variability on every spatial and timescale evolved, including extreme events like ``blowout'' of the central disk by winds, or rapid accretion of a large filament of gas (often from a non-linear spiral arm or eccentric mode in the outer disk, or from self-gravity-driven clumping where the sign of $B_{\phi}$ switches; see \citealt{hopkins:superzoom.imf}). But per HM25 and \S~\ref{sec:sims:tradeoffs}, it is hard to make robust quantitative statements about this variability: it is artificially ``stretched'' in time by the factors $a_{i}$, and the periods of largest dis-equilibrium are where the method is least reliable.

\subsection{What Is the Steady-State? Strongly-Magnetized Disks Throughout}
\label{sec:steady.state:mag}

\subsubsection{Hyper-Magnetized/Flux-Frozen, Cold, Toroidally-Magnetized Disks at Large $\dot{m}$ and $R$}
\label{sec:steady.state:mag:cold}

For most of the dynamic range in space, time, and $\dot{m}$ evolved, the steady-state remains a magnetically-dominated disk of the sort seen in \paperone\ and other simulations (\S~\ref{sec:intro}), broadly following analytic expectations \citep{hopkins:superzoom.analytic,begelman.pringle:2007.acc.disks.strong.toroidal.fields}. At $f_{\dot{M}}=0.03$ and $1$, $\beta \ll 1$ across essentially the entire disk (Fig.~\ref{fig:maps.pinwheel}, Table~\ref{tbl:runs}), rising toward unity only near the ROI, with magnetic pressure exceeding thermal (and comparable to or exceeding radiation) pressure throughout the inner disk. The turbulence is trans-\Alf{ic} and supersonic ($v_{\rm turb} \sim v_{A} \gg c_{s}$), roughly explaining the vertical support ($H \sim v_{\rm turb}/\Omega \sim v_{A}/\Omega$; Fig.~\ref{fig:height.breakdown}), and the fields are primarily coherent and toroidal ($| \langle B_{\phi} \rangle |  > \langle \delta B_{\phi}^{2} \rangle^{1/2}$ and $|B_{\phi}| \sim 3-10\, | B_{r,\,z} |$, outside of gaps from transient eccentric modes, warps, or large-scale $B_{\phi}$ sign changes), with weaker, turbulent poloidal fields ($|\langle B_{r,\,z} \rangle| \ll \langle \delta B_{r,\,z}^{2} \rangle^{1/2}$). As in \paperone, $H/R$ is a weak, increasing (flared) function of $R$; $B_{\phi}$ scales slightly steeper than $\propto r^{-1}$ (closer to $\sim r^{-5/4}$); $\rho(R) \propto R^{-(1.2-1.7)}$; and the thermal structure broadly follows the expectations of \citet{hopkins:superzoom.analytic} for a magnetically-dominated, externally-illuminated disk. The origins of the accretion stresses, field amplification by advection/flux-freezing, the roles of the different pressures in the vertical structure, and stability properties (including the lack of the magnetorotational instability and presence of Parker modes) are studied extensively in \paperone\ and \citet{hopkins:superzoom.disk,hopkins:superzoom.analytic}, so we do not repeat them here.

Although the inner-disk radiation energy density is broadly comparable to the magnetic at $\dot{m} \gtrsim 0.1$, most of the outer disk is optically-thin with a large scattering mean-free-path and weak midplane radiation pressure (\S~\ref{sec:rad}). At $\dot{m}\gg1$ the outer-disk midplane has a {\em downward}-pointing radiation flux (as expected for an externally-illuminated disk; \S~\ref{sec:rad:illum}), so radiation pressure cannot be holding it up.

\subsubsection{Magnetically Arrested, Hot Accretion at Small $\dot{m}$ and $R$}
\label{sec:steady.state:mag:hot}

The behavior at small radii $R \ll 1000\,R_{g}$ in our lowest-$\dot{m}$ ($f_{\dot{M}}=0.001$) simulation is markedly distinct. Like \citet{guo:2024.fluxfrozen.disks.lowmdot.ellipticals}, who simulated low-$\dot{m}$ flows from a hot ISM, we see an outer, ``cold'' magnetically-dominated disk from $\sim 500\,R_{g}$ to the circularization radius ($\sim 0.1\,$pc), with $\dot{M}_{\rm out} > \dot{M}_{\rm in}$ over this entire range (\S~\ref{sec:steady.state:acc}). Interior to this, the disk becomes dynamically and thermally ``hot'' (Fig.~\ref{fig:height.breakdown}): $H/R \sim 0.3-0.5$; volume-filling gas temperatures order-of-magnitude comparable to virial ($\gtrsim 10^{9}\,{\rm K}$); $\beta \sim 1$ (thermal and magnetic pressure comparable, both exceeding radiation); trans/sub-sonic turbulence; and an optically-thin ($\tau \lesssim 0.01$) central ``cavity'' in $\Sigma(R)$, with accretion dominated by clumps/filaments from the outer disk sinking through this region via magnetic flux eruptions and radial magnetic buoyancy instabilities. This is very much the classic ``Magnetically Arrested Disk'' (MAD) behavior \citep{bisnovatyi.kogan:1976.mad.disk,igumenshchev:2003.mad.sims,narayan:2003.mad.disk,tchekhovskoy:2011.mad.disk.jets}, with radiative properties resembling optically-thin, geometrically-thick advection-dominated accretion flows (ADAFs; \citealt{narayan.yi.95:adaf.lowmass.bhs,narayan.yi.95:adaf.self.similarity.outflows,yuan.narayan:2014.hot.accretion.review}): the radiative efficiency is low and the inner-disk emission is primarily non-thermal X-rays (\S~\ref{sec:rad}). As in \citet{white:2019.mad.disk.sims} and \citet{guo:2024.fluxfrozen.disks.lowmdot.ellipticals}, $B_{\phi}$ remains modestly dominant at small radii ($|B_{\phi}| \sim 1-3\,|B_{r,\,z}|$) but becomes less globally coherent (more MRI-like, $|\langle B_{\phi} \rangle | < \langle \delta B_{\phi}^{2} \rangle^{1/2}$, often flipping sign in the midplane with a more dipolar/odd-symmetry polarity than the more quadrupolar outer disk; see \citealt{kaaz:2024.hamr.forged.fire.zoom.to.grmhd.magnetized.disks}), while $B_{z}$ and $B_{r}$ become more coherent ($|\langle B_{r,\,z} \rangle | \sim \langle \delta B_{\phi}^{2} \rangle^{1/2}$).

Some details of this MAD regime are probably incorrect here: lacking full GR, there is no proper Blandford-Znajek jet, and a Newtonian sink boundary cannot fully capture the ``pileup'' of poloidal flux at the horizon \citep[see][]{kaaz:2024.hamr.forged.fire.zoom.to.grmhd.magnetized.disks}, so the coherent poloidal field at $\lesssim 10-100\,R_{g}$ is probably under-estimated. A detailed study of this low-$\dot{m}$ state, how the state change occurs, and the relevant critical $\dot{m}$ will be the subject of dedicated future work (Cho et al., in prep.).

\subsection{Accretion Rate Scaling With Radius: A Transition at Sufficiently Subcritical $\dot{M}$}
\label{sec:steady.state:acc}

\begin{figure*}
	\centering\includegraphics[width=0.49\textwidth]{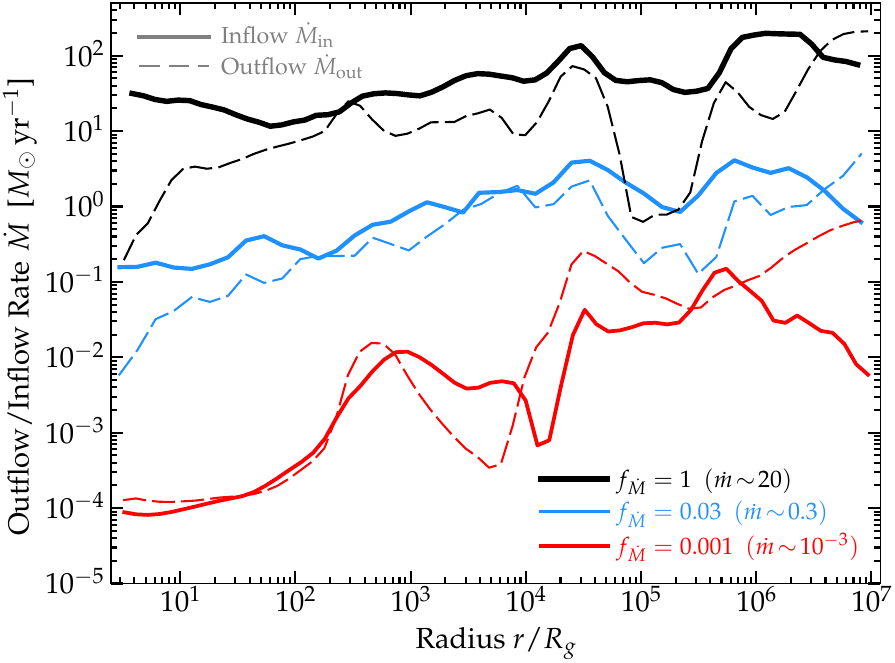}\includegraphics[width=0.49\textwidth]{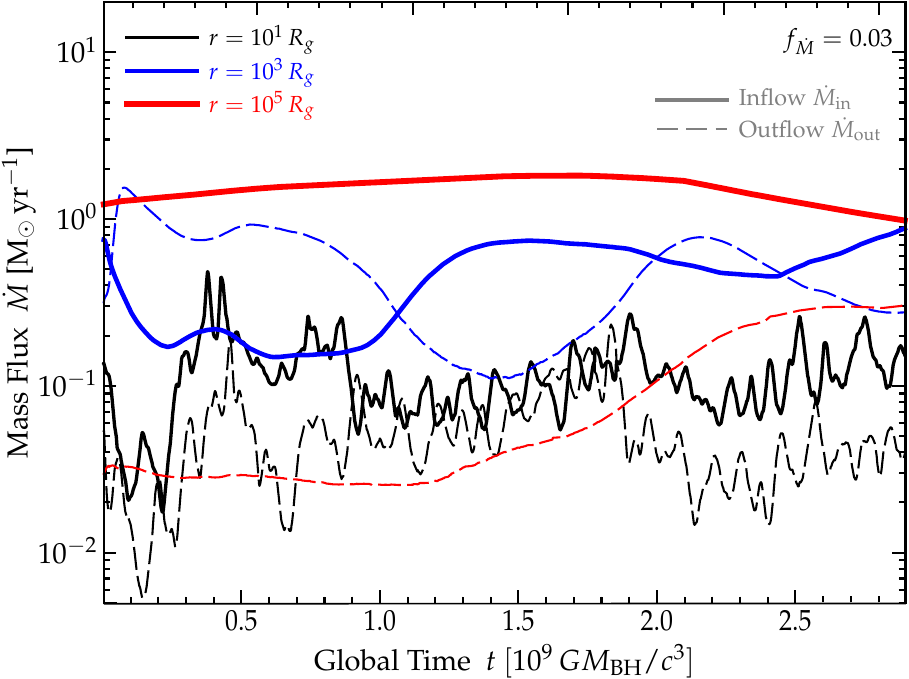}
	\caption{Inflow \&\ outflow rates vs.\ radius ({\em left}) and time ({\em right}). We compare the three accretion-rate cases $f_{\dot{M}}=(1,0.03,0.001)$ corresponding to $\dot{m}\sim(20,0.3,0.001)$, with $\dot{m}\sim0.3$ ($\dot{M} \sim 0.1\,{\rm M}_{\odot}\,{\rm yr^{-1}}$) our fiducial case. Profiles average over the last $\sim 10\%$ of the simulation runtime. Broadly, $\dot{M}_{\rm out} \sim \dot{M}_{\rm in}$. At high $\dot{m}$, $\dot{M}$ is only weakly $r$-dependent; this steepens at $\dot{M} \lesssim 0.01$. We evolve to $\sim 10^{9.5}\,G M_{\rm BH}/c^{3}$; the system is in quasi-steady-state at $r \lesssim 10^{5}\,R_{g}$ on these scales. Variability is more-rapid at small-$r$, but the variability timescales are biased by our dilation scheme: de-dilated timescales for $\mathcal{O}(1)$ variability in $\dot{M}$ are $\mathcal{O}(10\,[r^{3}/G M_{\rm BH}]^{1/2})$.\label{fig:mdot}}
\end{figure*}

At high $\dot{m}$ ($f_{\dot{M}} = 0.03,\,1$), $\dot{M}_{\rm in}$ decreases somewhat from the ROI to the ISCO (Fig.~\ref{fig:mdot}), but (1) only by a factor of $\sim 10$ at most in the fiducial $f_{\dot{M}}=0.03$ case (and less at $f_{\dot{M}}=1$), i.e.\ $\dot{M}_{\rm in} \propto R^{0.1-0.2}$ given the very large ($\sim 10^{7}$) dynamic range in $R$; and (2) much of the decrease occurs at intermediate radii $100\, R_{g} \ll R \ll R_{\rm ROI}$, corresponding to the BLR and inner torus, where strong radiation-pressure driven winds are launched (\S~\ref{sec:outflow:high.mdot:rad}). Within the central few hundred (few $10^{4}$) $R_{g}$ for $f_{\dot{M}}=0.03$ ($=1$), $\dot{M}(R)$ is remarkably constant.

At $f_{\dot{M}} = 0.001$, by contrast, $\dot{M}_{\rm in}$ decreases steadily inwards. Individual radii still fluctuate dramatically, but on average this is consistent with simulations of low-luminosity AGN accreting from hot (virialized) gaseous halos at $\dot{m} \ll 0.01$ \citep{guo:2022.superzoom.riaf.in.m87.nosf.nomhd.etc,guo:2024.fluxfrozen.disks.lowmdot.ellipticals,cho:2023.multiscale.accretion.sims.bondi.inflow.model}, which found $\dot{M}_{\rm in}(r) \propto r^{1/2}$. The other properties of the flow are also consistent with these quasi-spherical hot accretion flow models: modestly super-sonic and \Alf{ic} turbulence with $\beta \sim 1$, and $\rho \propto r^{-1}$ (``1-law'' accretion, as opposed to the Bondi-like $3/2$, consistent with a turbulent flow with roughly constant momentum flux; \citealt{narayan.yi.95:adaf.self.similarity.outflows,quataert.gruzinov:2000.cdaf,yuan.narayan:2014.hot.accretion.review}).

\begin{figure*}
	\centering\includegraphics[width=0.485\textwidth]{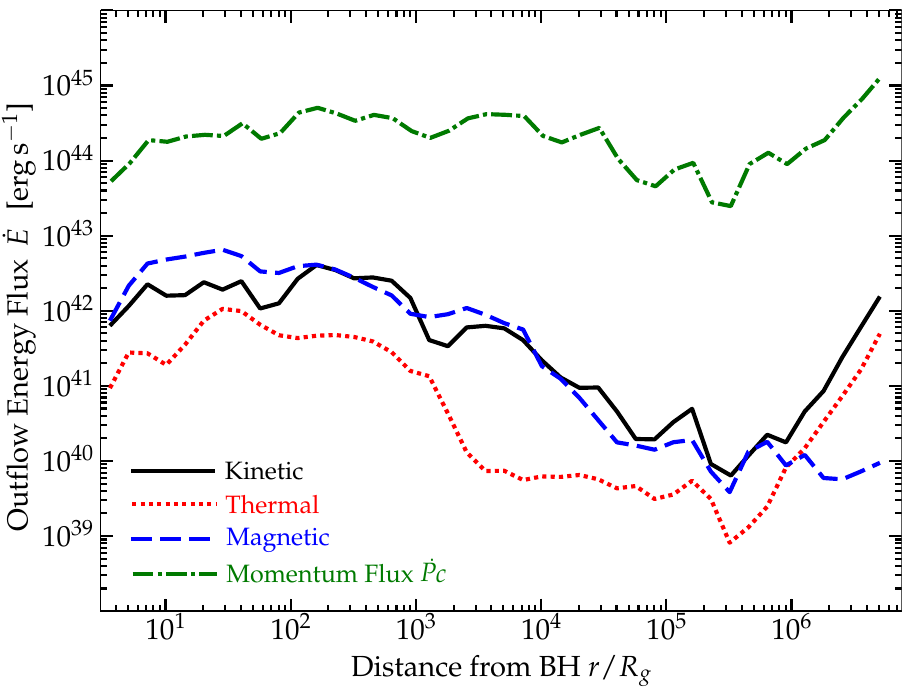}\includegraphics[width=0.49\textwidth]{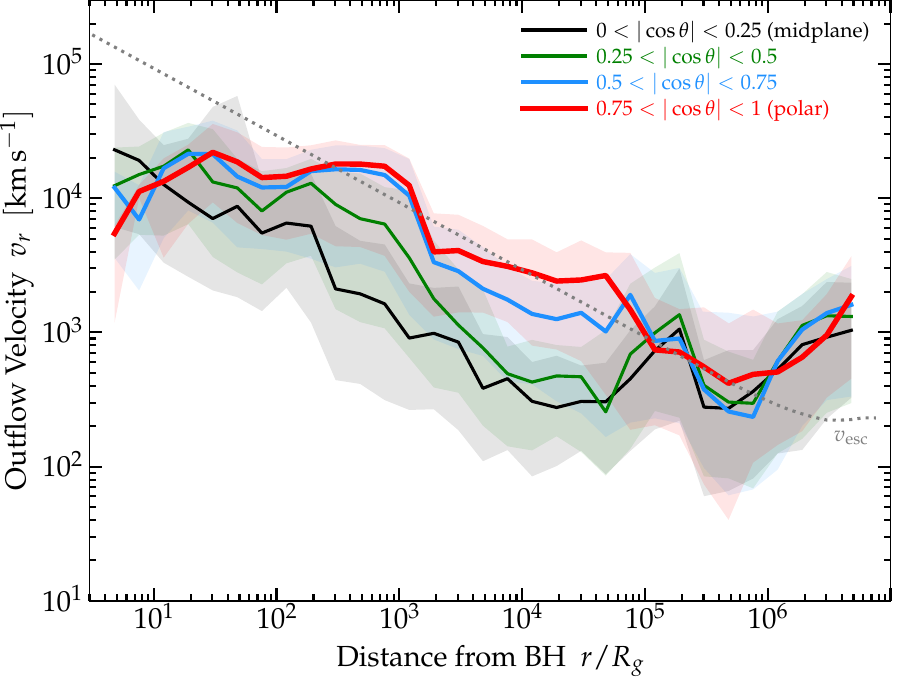}
	\caption{Outflow energy+momentum fluxes ({\em left}) and velocities ({\em right}) versus radius and polar angle, for the fiducial $\dot{m}\sim 0.3$ case. As Fig.~\ref{fig:mdot} we plot radial profiles in shells at the end of the simulation. We compare the kinetic, thermal, and magnetic energy flux of outflowing material ($v_{r} > 0$), and the total outflow momentum flux (multiplied by $c$ to compare units), and the velocities shown are binned by polar angle $\cos{\theta}$ with the kinetic-energy-weighted median ({\em line}) and inter-quartile range ({\em shaded}) shown (vs.\ escape velocity $v_{\rm esc}(r)$). The outflows feature $\dot{E}_{\rm kin} \sim \dot{E}_{\rm B} \sim 10\,\dot{E}_{\rm th}$ over their launching radii, with median $v_{r}\sim v_{\rm esc}$ over a broad bicone covering half the sky. The decrease of $\dot{E}$ and $v_{r}$ owes to entrainment and new material being accelerated into outflow: $\dot{P}$ appears to be closer-to-constant with $r$ than $\dot{E}$.\label{fig:outflow.Ev}}
\end{figure*}

\begin{figure*}
	\centering\includegraphics[width=0.485\textwidth]{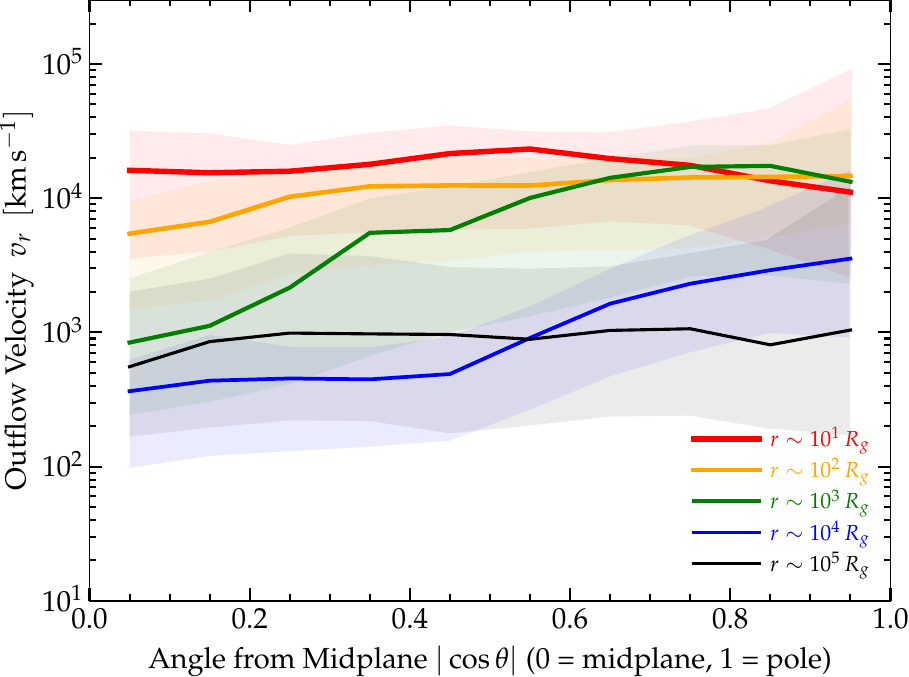}\includegraphics[width=0.49\textwidth]{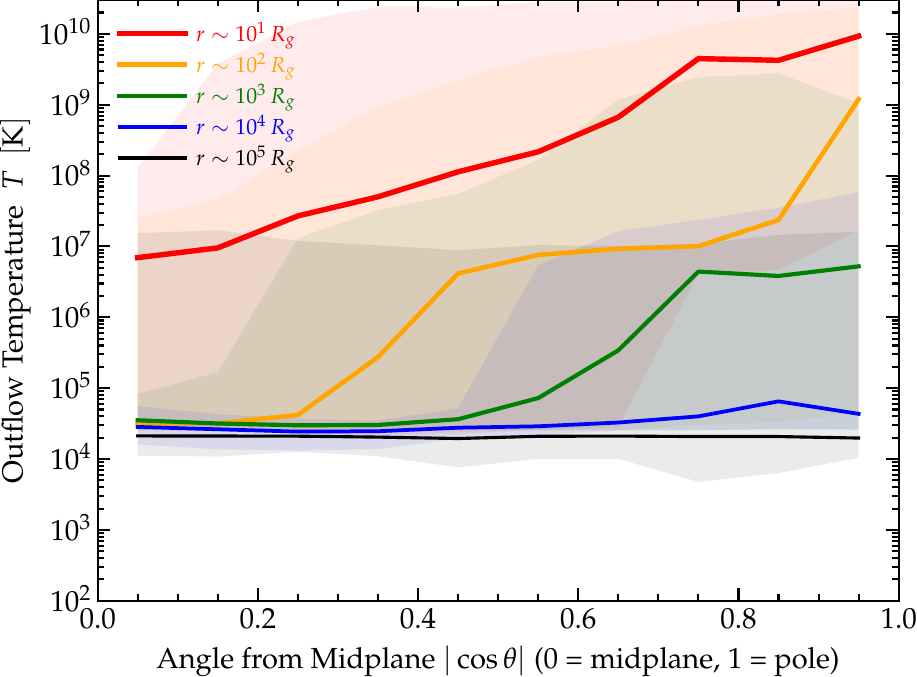}
	\caption{Outflow velocity $v_{r}$ and  temperature $T$ versus polar angle $\cos{\theta}$ and radius (same $\dot{m}\sim0.3$ case and time as Fig.~\ref{fig:outflow.Ev}). Kinetic-energy weighted median ({\em lines}) and $1-99\%$ range ({\em shaded}) shown. Polar outflows tend to be hotter and faster, though the latter trend is more visible in the high-$v_{r}$ tail of the velocity distribution than in the median. Outflow $v_{r}$ ranges from $\sim 100\,{\rm km\,s^{-1}}$ for slower outflows at $\gg 10^{4}\,R_{g}$ to $\sim 10^{4}\,{\rm km\,s^{-1}}$ for the tail of escaping outflows, with temperatures $T \sim 10^{4}-10^{5}\,$K at these radii, and up to $\sim 0.1-0.3\,c$ for hot ($T \sim  10-100\,$keV) polar outflows at $r \ll 100\,R_{g}$ (recall we do not have GRMHD, this is with no jet).\label{fig:outflow.Tv.angle}}
\end{figure*}

\begin{figure*}
	\centering
	\includegraphics[width=0.99\textwidth]{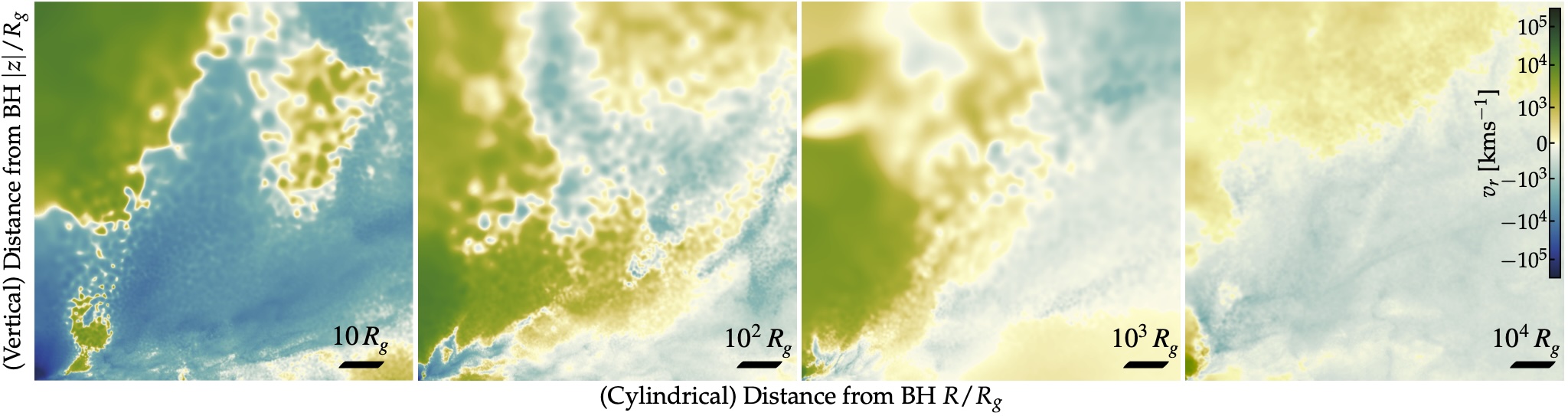}\\
	\hspace{10pt}\includegraphics[width=0.97\textwidth]{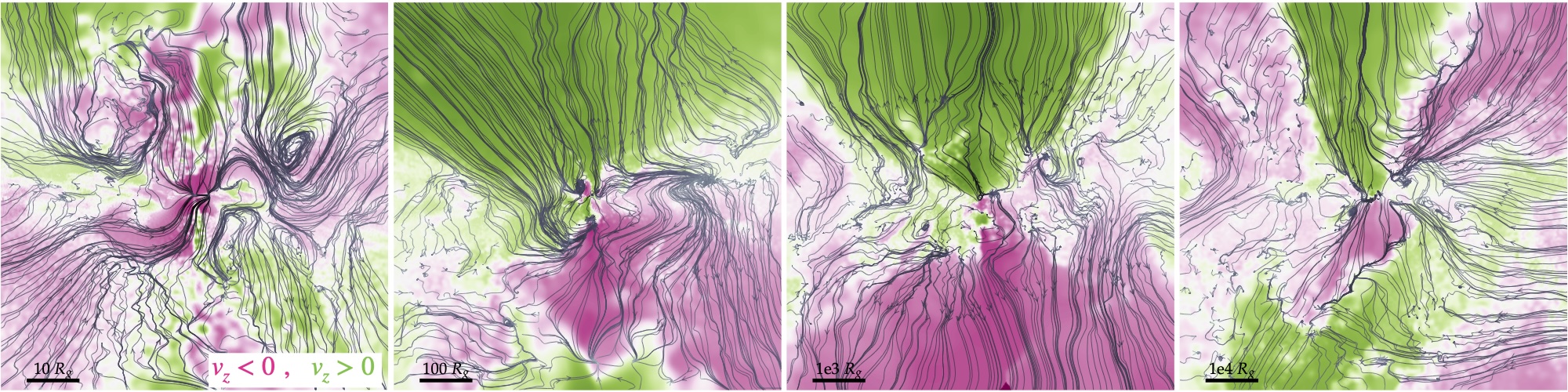}
	\caption{Example images showing small-scale outflow velocity structure. {\em Top:} We take $R-z$ slices (same $\dot{m}\sim0.3$ case and final time as Fig.~\ref{fig:outflow.Ev}), through the $y$-axis, and plot for every pixel the interpolated radial velocity $v_{r}$ of the nearest cell center. 
	Blue is inflow, green is outflow, with velocities on a log-scale from $\sim100\,{\rm km\,s^{-1}}$ to $\pm c$. Per Fig.~\ref{fig:outflow.Tv.angle}, outflows are faster at smaller-$r$ and more polar angles, but there is considerable substructure at all radii and positions. Lower angles correspond to the disk and its thick, flared layers. {\em Bottom:} $x-z$ slice images (BH at center) showing the in-plane velocity flowlines, colored by $v_{z}$. \label{fig:profiles.outflow.velocity}}
\end{figure*}

\section{Outflows \&\ Jets at Large Radii}
\label{sec:outflow}

We see strong outflows in the simulations. We show the mass ($\dot{M}_{\rm out}$; Fig.~\ref{fig:mdot}), momentum ($\dot{P}_{\rm out}$), and kinetic/magnetic/thermal ($\dot{E}_{\rm out}^{\rm kin}$, $\dot{E}_{\rm out}^{\rm B}$, $\dot{E}_{\rm out}^{\rm th}$) energy fluxes (Fig.~\ref{fig:outflow.Ev}) of outflowing gas, and its velocity and temperature structure versus radius and angle (Figs.~\ref{fig:outflow.Tv.angle}--\ref{fig:profiles.outflow.velocity}); Fig.~\ref{fig:profiles.rad} shows the outward radiative acceleration at different angles and radii.

Compared to inflowing gas at similar radii, the outflows tend to be (1) more concentrated towards polar regions; (2) lower-density; (3) hotter (often $\sim 10^{9}\,$K near the outflow ``base,'' with more cool gas mixed in at larger radii; Fig.~\ref{fig:outflow.Tv.angle}), with higher thermal energy flux ($|\dot{E}_{\rm out}^{\rm th}| > |\dot{E}_{\rm in}^{\rm th}|$); (4) faster-moving (in both $|{\bf v}|$ and $|v_{r}|$, with $|\dot{E}_{\rm out}^{\rm kin}| > |\dot{E}_{\rm in}^{\rm kin}|$); (5) higher-$\beta$ (typically $\beta > 1$, sometimes $\gtrsim 100$, owing to the high temperatures); and (6) the channels along which UV/soft-X-ray ($T_{\rm rad} \sim 10^{5}-10^{7}\,$K) radiation efficiently escapes (owing to their lower densities and absorption opacities).

Geometrically, the broad-angle outflows tend to start ``just above'' the disk, launching at intermediate radii $\sim$\,10's to 100's of $R_{g}$ before being collimated by the thick, flared disk into more bipolar structures at larger radii (Fig.~\ref{fig:profiles.outflow.velocity}). At even larger radii they can become more disordered again, owing to interactions with inflows and large-scale warps/changes in the orientation of the outer disk (see \paperone\ and \citealt{hopkins:superzoom.disk}). The outflows change qualitatively from the high-$\dot{m} \gg 0.01$ to low-$\dot{m} \lesssim 0.01$ regime, consistent with the changes in the accretion flow above, so we discuss each regime in turn.

\subsection{High-$\dot{m}$: Strong Bipolar Radiation-Pressure Driven Outflows Predominate}
\label{sec:outflow:high.mdot}

We first consider the high-$\dot{m}$ ($\dot{m} \gtrsim 0.05$, i.e.\ $f_{\dot{M}} = 0.03,\,1$) cases.

\subsubsection{Weak Magnetocentrifugal Winds}
\label{sec:outflow:high.mdot:Bwinds}

At high-$\dot{m}$, we see no evidence for strong, large-scale magnetocentrifugal (\citealt{blandfordpayne:mhd.jets} or \citealt{koniglkartje:disk.winds}-type) winds. (1) Directly measuring the radial MHD acceleration $a_{r,\,{\rm MHD}} \equiv {\bf a}_{\bf MHD} \cdot \hat{\bf r}$ on each gas cell, it is almost always sub-dominant to gravity and to radiation forces in the outflow-launching regions. (2) The simulations in \citet{hopkins:superzoom.overview}, which had the same magnetic/gas configuration outside the inner disk (so should have similar magnetocentrifugal winds) but no radiation/illumination from the inner disk, showed polar inflow, not outflow. (3) As discussed in \citet{hopkins:superzoom.disk}, the column densities and ram pressures at the top of the disk are, at high $\dot{m}$, well above the critical values where idealized simulations predict magnetocentrifugal winds cannot launch and are instead ``trapped'' into a fountain flow at heights $\sim H$ \citep{ouyed:1999.numerical.sims.mhd.winds.vs.massloading,anderson:2005.mass.loading.mhd.winds}. (4) The field-line and velocity-streamline structure is not that predicted for magnetocentrifugal winds: the poloidal field is associated with large turbulent-like loops with coherence lengths up to $\sim H$ \citep[see \paperone\ and][]{kaaz:2024.hamr.forged.fire.zoom.to.grmhd.magnetized.disks}, lacking the large-scale coherence to $|z|\rightarrow\infty$ these outflows require, and the velocity streamlines show acceleration off the surface of the disk radially away from the center (Fig.~\ref{fig:profiles.outflow.velocity}), not a tower-like morphology.

\subsubsection{Weak Jets (at Equilibrium Spin)}
\label{sec:outflow:high.mdot:jets}

In a subset of our runs, we include an explicit sub-grid jet injection motivated by the low-spin equilibrium found in \citet{Ricarte2023,Lowell2024, Lowell2025}: at high $\dot{m}$, chaotic accretion of clumpy, magnetically-dominated gas efficiently spins down the BH, driving the spin toward modest $a \sim 0.1-0.3$ (rather than the often-assumed $a \sim 1$). The Blandford-Znajek jet power (\S~\ref{sec:subgrid:scalings}) is then modest, $\dot{E}_{j} \sim a^{2} (H/R)^{2}_{\rm ISCO} \dot{M}_{\rm BH} c^{2} \sim 10^{-3}-10^{-2}\,\dot{M}_{\rm BH}\,c^{2}$, comparable to or smaller than the radiative luminosity. More importantly, the jets must propagate through the dense, thick disk atmosphere and polar regions, which at high $\dot{m}$ have ram pressures orders of magnitude larger than the jet energy flux, so they are rapidly decelerated and disrupted within $\lesssim 100\,R_{g}$, heating the coronal gas rather than producing coherent large-scale outflows. We see this directly: jets injected at the inner boundary modestly perturb the thermal state of their immediate surroundings but produce no discernible large-scale collimated outflow. The dominant outflow channel at high-$\dot{m}$ is overwhelmingly radiation pressure.

\subsubsection{Strong Radiation Pressure-Driven Outflows}
\label{sec:outflow:high.mdot:rad}

\begin{figure*}
	\centering\includegraphics[width=0.32\textwidth]{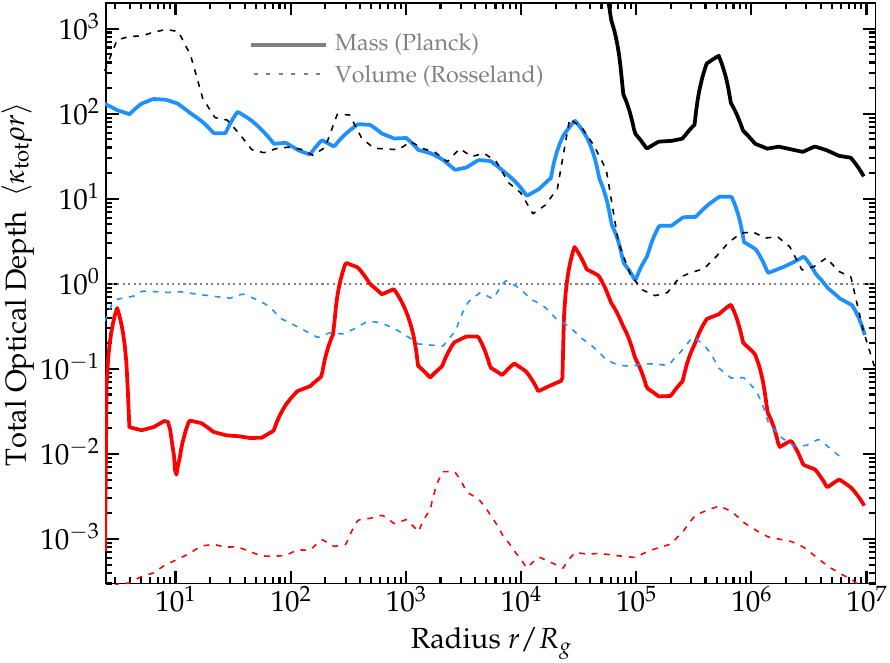}\includegraphics[width=0.32\textwidth]{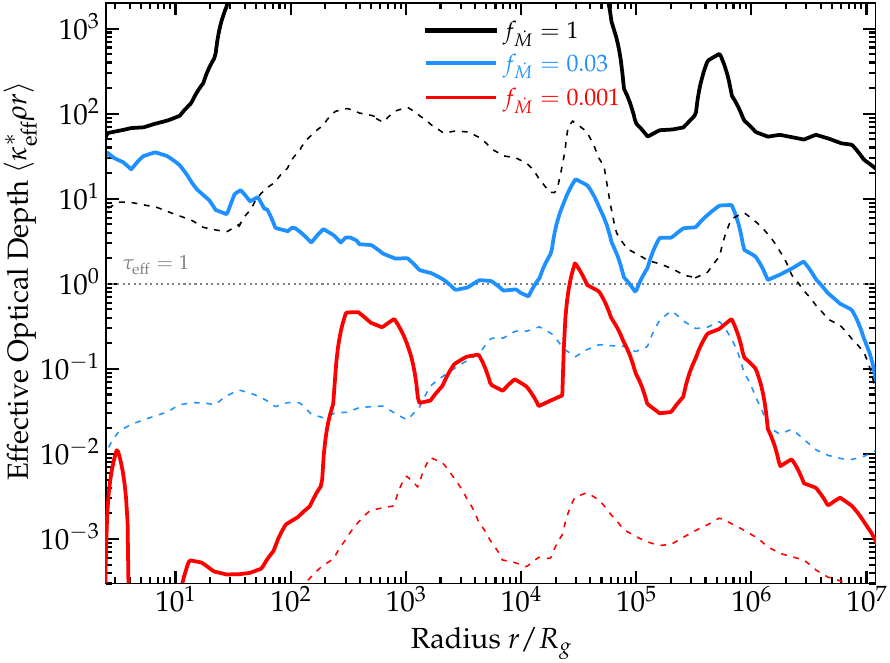}\includegraphics[width=0.32\textwidth]{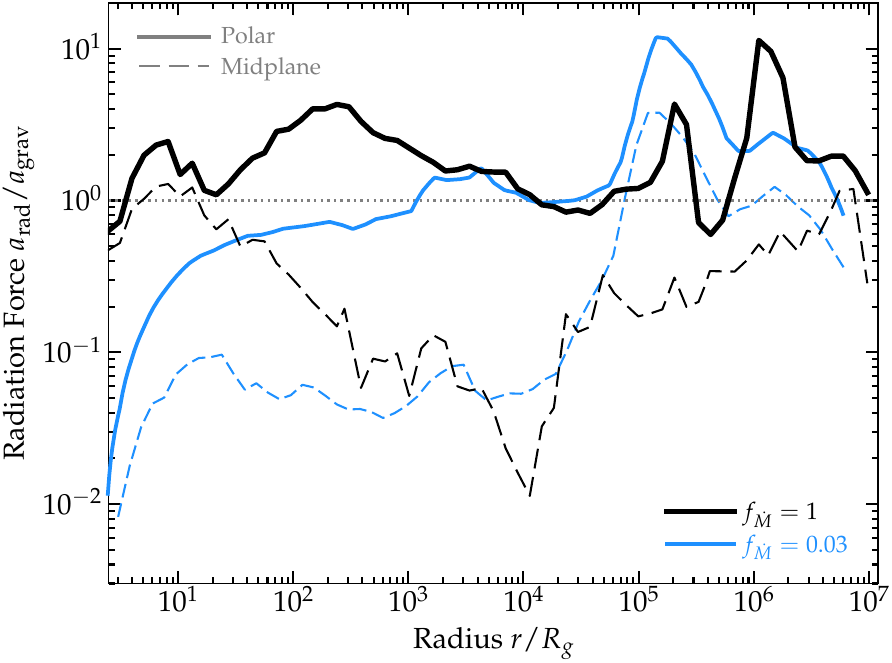}
	\caption{Radiative properties versus radius, for the three runs (as Fig.~\ref{fig:lum}, at the final simulation time). {\em Left:} ``Total optical depth'' seen by a photon packet depends on the entire system; here we plot a simple proxy of the optical depth through each radius, $\langle \kappa \rho r\rangle$, either using a mass-weighted Planck mean (more relevant for disk emission) or a volume-weighted Rosseland mean (for polar escape), using the on-the-fly radiation spectrum at each point to calculate $\kappa$. We compare total $\kappa_{\rm tot}$ and effective $\kappa_{\rm eff} \equiv \sqrt{\kappa_{\rm tot} \kappa_{\rm abs}}$. The disk at $\ll 10^{3}\,R_{g}$ has $\tau_{\rm eff} \gg 1$, giving rise to greybody (big-blue bump) emission, at high $\dot{m}$, but at $\dot{m} \ll 0.01$ the entire disk is optically thin. Spikes at larger $r$ correspond to opacity jumps (e.g.\ dust formation). {\em Right:} Ratio of radial outward radiation-to-gravitational acceleration $a_{\rm rad}/a_{\rm grav}$, volume-averaged in a polar wedge ($|\cos\theta|>0.75$) versus mass-weighted in a midplane wedge ($|\cos\theta|<0.25$), for high-$\dot{m}$ (the ratio is $\ll1$ everywhere for $\dot{m} \sim 0.001$). The biconical outflows are driven by radiation pressure, while accretion through the midplane is weakly modified except at transient regions where opacity changes or disk geometry changes appear.\label{fig:profiles.rad}}
\end{figure*}

The outflows at high-$\dot{m}$ are clearly radiation-pressure driven. This is not surprising at $\dot{m} \gtrsim 0.1$, but we confirm that the radiative acceleration $a_{\rm rad} \equiv {\bf a}_{\rm rad} \cdot \hat{\bf r}$ is the dominant force on the launched material and locally exceeds gravity (Fig.~\ref{fig:profiles.rad}), that the outflows trace the radiation flow lines (Fig.~\ref{fig:maps.rad.streamlines}), and that they accelerate to the characteristic speeds expected for radiation-driven flows.

There is some outflow at all scales, but it is clearly not a scale-free process, with preferred ``launching radii'' (Figs.~\ref{fig:mdot}--\ref{fig:outflow.Ev}), most obvious in the radiative acceleration, with strong launching a bit outside $\sim 100\,R_{g}$. At smaller radii, the outflows have small mass-loading and are mostly tenuous, hot ($\gg 10^{5}\,$K) gas in a ``flat'' layer above the midplane without strong polar collimation: with $\beta_{\rm rad} \sim 1$, this is more akin to an evaporating thick-disk atmosphere, as predicted for quasi-hydrostatic but inhomogeneous/turbulent radiation-pressure supported atmospheres \citep{jiang:2019.superedd.sims.smbh.prad.pmag.modest.outflows,utsumi:2022.superedd.rmhd.sims.accretion.mag.comparable.rad}. In the inner disk ($\lesssim 100\,R_{g}$) electron scattering dominates the opacity, but at somewhat larger radii the effective temperature of gas and radiation drops to $\sim 10^{4}-10^{5}\,$K and the absorption opacity of at least some of the gas above the disk rises dramatically (Fig.~\ref{fig:profiles.rad}; e.g.\ Kramers opacity $\propto \rho\,T^{-7/2}$, plus non-linearly increasing metal-line and bound-free contributions, with absorption eventually exceeding scattering by orders of magnitude in the outer disk; \citealt{lenzuni:1991.opacities,semenov:2003.dust.opacities}). Our multi-group treatment captures these transitions self-consistently, which is essential for predicting where outflows are launched. In the polar wedge ($|\cos\theta|>0.75$) $a_{\rm rad} \gtrsim a_{\rm grav}$ over $\sim30-300\,R_{g}$ (and again near the dusty launching zone at $\sim10^{5}\,R_{g}$), while in the midplane ($|\cos\theta|<0.25$) $a_{\rm rad}\ll a_{\rm grav}$: the midplane remains a well-shielded, externally-irradiated layer (\S~\ref{sec:rad:illum}), and the outflow is launched preferentially at high latitude. Since magnetic support keeps the disk thick at these radii, there is plenty of material at large heights to accelerate, and both $\dot{M}_{\rm out}$ (Fig.~\ref{fig:mdot}) and $\dot{P}_{\rm out}$ (Fig.~\ref{fig:outflow.Ev}) jump. New material continues to be accelerated into the outflow from the surface layers of the flared disk at larger radii (Fig.~\ref{fig:profiles.outflow.velocity}), so $\dot{M}_{\rm out}(r)$ continues to increase, until the ambient medium becomes uniformly dusty and the radiation is rapidly downgraded to infrared $T_{\rm rad}$, so the opacities and radiative accelerations drop. Over these radii the disk structure shapes the outflows into more coherent broad-angle bipolar flows (Fig.~\ref{fig:profiles.outflow.velocity}).

In terms of ``loading factors,''\footnote{As in \paperone, we caution that many of the oscillations in $\dot{M}_{\rm in}$ and $\dot{M}_{\rm out}$ at very large radii, where the simulation is only evolved for a few dynamical times, owe not to true outflows or inflows but to eccentric/lopsided/$m=1$ modes in the disk, which our simple spherical-shell definition of $\dot{M}_{\rm in/out}$ does not account for.} $\dot{M}_{\rm out}$ at the largest radii of the outflows is a few to ten times the inflow rate onto the BH (Fig.~\ref{fig:mdot}), consistent with the difference between the BH accretion rate and that through the ROI. This is not negligible, but it is not a major qualitative change to our intuition, nor sufficient to eliminate highly super-critical accretion: the corresponding $\dot{M}(r) \propto r^{\alpha}$ with $0.1\lesssim \alpha \lesssim 0.2$ (\S~\ref{sec:steady.state:acc}) is a very weak ``correction'' to analytic disk models like \citet{hopkins:superzoom.analytic,begelman.pringle:2007.acc.disks.strong.toroidal.fields}. In momentum, the sub-critical ($\dot{m} \sim 0.1-1$) cases show $\dot{P}_{\rm out} \approx L/c$ (with $L$ the escaping luminosity; Fig.~\ref{fig:lum}), while the most super-critical times/cases, reaching transient $\dot{m} \sim 20-50$, show $\dot{P}_{\rm out} \approx 10\,L/c$, with more like $\dot{P}_{\rm out} \propto \dot{m} L/c$ when $1 \lesssim \dot{m} \lesssim 10$ (Fig.~\ref{fig:profiles.outflow.loadings}), consistent with basic expectations for radiation-pressure-driven outflows \citep{murray:momentum.winds,thompson:rad.pressure,proga:disk.winds.2000b}. Despite local regions with effective optical depth $\gg 1$ (Fig.~\ref{fig:profiles.rad}), at sub-critical Eddington ratios there are escape channels for the radiation, which can also be downgraded to lower-opacity frequencies, but an $\mathcal{O}(1)$ fraction of the sky is covered by the thick disk and optically-thick material, so the outflow momentum is comparable to the single-scattering expectation. As $\dot{m}$ exceeds unity, the optical depth of the entire sky seen by the inner disk increases $\propto \rho \propto \dot{m}$, so multiple-scattering gives $\dot{P} \sim \tau_{\rm eff} L/c \sim \dot{m} L/c$, but this saturates at $\dot{P} \sim 10\,L/c$ (at $\dot{m} \gtrsim \dot{m}_{\rm sat} \sim 10$), consistent with idealized, nonlinear simulations of radiation-pressure driven outflows \citep{thompson:rad.pressure,hopkins:2019.grudic.photon.momentum.rad.pressure.coupling,kuiper:2012.rad.pressure.outflow.vs.rt.method}.

In terms of energy flux (Fig.~\ref{fig:outflow.Ev}), the outflows are kinetic-dominated with $\dot{E}_{\rm K}^{\rm out} \sim (1/2) \dot{P}^{2}_{\rm out}/\dot{M}_{\rm out}$. Near-ISCO, the magnetic loading $\dot{E}_{\rm B}^{\rm out} \sim \dot{E}_{\rm K}^{\rm out}$. In the lower-$\dot{m}\sim (0.3,\,0.001)$ runs, $\dot{E}_{\rm B}^{\rm out}$ drops more rapidly ($\propto 1/r$,  consistent with simple flux-freezing expectations $|{\bf B}|\propto r^{-2}$ in an isotropic outflow) than $\dot{E}_{\rm K}^{\rm out}$ in the launching zone and becomes smaller ($\lesssim 0.1\,\dot{E}_{\rm K}^{\rm out}$) at larger radii. The thermal-loading is small throughout, $\dot{E}_{\rm th}^{\rm out} \sim (0.01-0.1)\,\dot{E}_{\rm K}^{\rm out}$ (toward the lower end at $\dot{m} > 1$). Since the outflow itself is optically-thin and non-relativistic, there is no meaningful ``radiation-loading.''

Figs.~\ref{fig:outflow.Ev}--\ref{fig:outflow.Tv.angle} show the velocities versus radius and polar angle in detail: characteristic velocities span $\sim 10^{3}-10^{5}\,{\rm km\,s^{-1}}$, more like $\sim 5000-30,000\,{\rm km\,s^{-1}}$ at most angles around the putative BLR, similar to typical quasar outflows from these radii \citep{laor:warm.absorber,crenshaw:nlr,dunn:agn.fb.from.strong.outflows,faucher-giguere:2012.felobal.model}.

\subsection{Low-$\dot{m}$: Slow Mass-Loaded Convective Outflows and Fast Jets}
\label{sec:outflow:low.mdot}

At low-$\dot{m} \ll 0.01$ (our $f_{\dot{M}}=0.001$ case), along with the change in accretion state and radiative properties, the outflows change qualitatively.

\subsubsection{Lack of Radiation-Pressure Driven Outflows}
\label{sec:outflow:low.mdot:rad}

Unsurprisingly, with the flow radiatively-inefficient so $L \sim 10^{-4}\,L_{\rm Edd}$, radiation-pressure driven outflows cease. The flow is optically-thin, so the ratio of radiative-to-gravitational force is $\sim (L/L_{\rm Edd})\,(\kappa/\kappa_{\rm es})$, and since the spectrum lacks a ``big blue bump,'' $\kappa$ (even in the dusty torus) is never much larger than $\kappa_{\rm es}$. We confirm this directly in the measured opacities and accelerations (Fig.~\ref{fig:profiles.rad}).

\subsubsection{Turbulent/Convective Outflows}
\label{sec:outflow:low.mdot:winds}

Even excluding jets, we still see outflows at $\dot{m} \ll 1$, with $\dot{M}_{\rm out} \sim \dot{M}_{\rm in}$ as needed to explain $\dot{M}_{\rm in} \propto r^{1/2}$ or $\dot{M}_{\rm in}^{\rm BH} \ll \dot{M}_{\rm in}^{\rm ROI}$. These outflows are hot ($c_{s} \sim V_{c}$), low-density, not strongly magnetized ($\beta \gtrsim 1$), and spatially and temporally intermittent, with coherence lengths/times $\sim r$ and $\sim t_{\rm orbit}$ at each radius and frequent sign flips in $v_{r}$. They are often quasi-spherical, occupying the entire volume at a given radius with alternating outflow and inflow (the ``outflow'' regions corresponding to ``holes'' in the accretion flow, sometimes punctuated by sinking clumps), rather than a vertically-stratified ``disk+outflow'' morphology; even where the latter is present, the ``opening angle'' is almost the entire sky excepting the portion shielded by the cold disk, as expected for outflows following a path of ``least resistance.'' This is similar to general expectations for turbulence/convection/buoyancy-driven outflows in quasi-spherical accretion flows \citep{narayan.yi.95:adaf.self.similarity.outflows,quataert.gruzinov:2000.cdaf,yuan.narayan:2014.hot.accretion.review}: a quasi-spherical, near-hydrostatic ($|\langle \dot{M} \rangle | \ll M_{\rm gas} \Omega$) flow in a Keplerian potential generically violates the Schwarzschild criterion for convective stability, driving $\dot{M}_{\rm out} \sim \dot{M}_{\rm in}$ with most of the integrated outflow coming from the largest radii (here the ROI), at characteristic velocities of order a couple $v_{\rm K}(R_{\rm ROI})$ (comparable to the escape velocity) that are trans or modestly super-sonic/\Alf{ic}. Compared to the radiation-pressure-driven winds launched from smaller radii at high-$\dot{m}$, these are therefore comparably ``slow.''

\subsubsection{Jet Escape}
\label{sec:outflow:low.mdot:jets}

In sharp contrast to high $\dot{m}$, at $\dot{m} \ll 0.01$ the conditions become far more favorable for jet escape to large radii. We stress that we do {\em not} self-consistently generate or propagate a relativistic jet (we lack GRMHD, and jets enter only via the crude sub-grid injection of \S~\ref{sec:outflow:high.mdot:jets}); the following estimates use the {\em ambient} conditions ($n$, $u_{B}$, $u_{\rm rad}$) measured in the simulations. The gas density in the central $\sim 1-100\,R_{g}$ is a factor $\sim 10^{5}-10^{6}$ lower than at the highest $\dot{m}$ ($n \sim 10^{8}$ vs.\ $\sim 10^{14}\,{\rm cm^{-3}}$ at $100\,R_{g}$, a highly super-linear decrease with $\dot{m}$), so entrainment is far weaker: at high $\dot{m}$ the dense polar material decelerates a jet to non-relativistic speeds within $\sim 10-100\,R_{g}$, while at low $\dot{m}$ the evacuated polar ``funnel'' allows it to remain relativistic out to $\gtrsim 10^{3}-10^{4}\,R_{g}$, consistent with the classic picture of powerful radio jets escaping primarily in low-luminosity systems \citep{yuan.narayan:2014.hot.accretion.review}.

Radiative losses tell the same story. The dominant loss channels for relativistic leptons are inverse Compton (IC) scattering off $u_{\rm rad}$ and synchrotron ($u_{B}$). At low $\dot{m}$, $u_{B} \gtrsim 10-100\,u_{\rm rad}$ throughout the inner regions (including the polar regions), with $u_{B} \sim 10^{4}-10^{5}/x^{2}\,{\rm erg\,cm^{-3}}$ ($x \equiv r/R_{g}$), while at high $\dot{m}$, $u_{B} \sim u_{\rm rad}$ in the midplane and $u_{B} < u_{\rm rad}$ in the polar regions, with $u_{B} + u_{\rm rad} \sim 10^{10}/x^{2}\,{\rm erg\,cm^{-3}}$. For a GeV lepton at $x = 10-100$, the loss timescale is then $\sim 10^{-4}-10^{-2}\,{\rm s}$ (IC-dominated) at high $\dot{m}$ versus $\sim 10^{4}\,{\rm s}$ (synchrotron-dominated) at low $\dot{m}$, compared to a light-crossing time $\sim 10^{4}\,m_{7}\,(x/100)\,{\rm s}$: at low $\dot{m}$ the jet can transport energy to large radii with synchrotron-dominated emission resembling traditional radio sources, while at high $\dot{m}$ the jet energy goes into coronal heating. Hadronic losses (timescale $\propto 1/n$) differ even more, $\sim 30-300\,{\rm s}$ at high $\dot{m}$ versus $\sim 3 \times 10^{7}\,{\rm s}$ at low $\dot{m}$. At $\gtrsim 10^{4}-10^{5}\,R_{g}$, stellar radiation from the dense nuclear cluster becomes non-negligible; this is unlikely to stop a relativistic jet, but could add entrainment and mass-loading at larger radii, which we defer to future work.

\begin{figure*}
	\centering\includegraphics[width=0.48\textwidth]{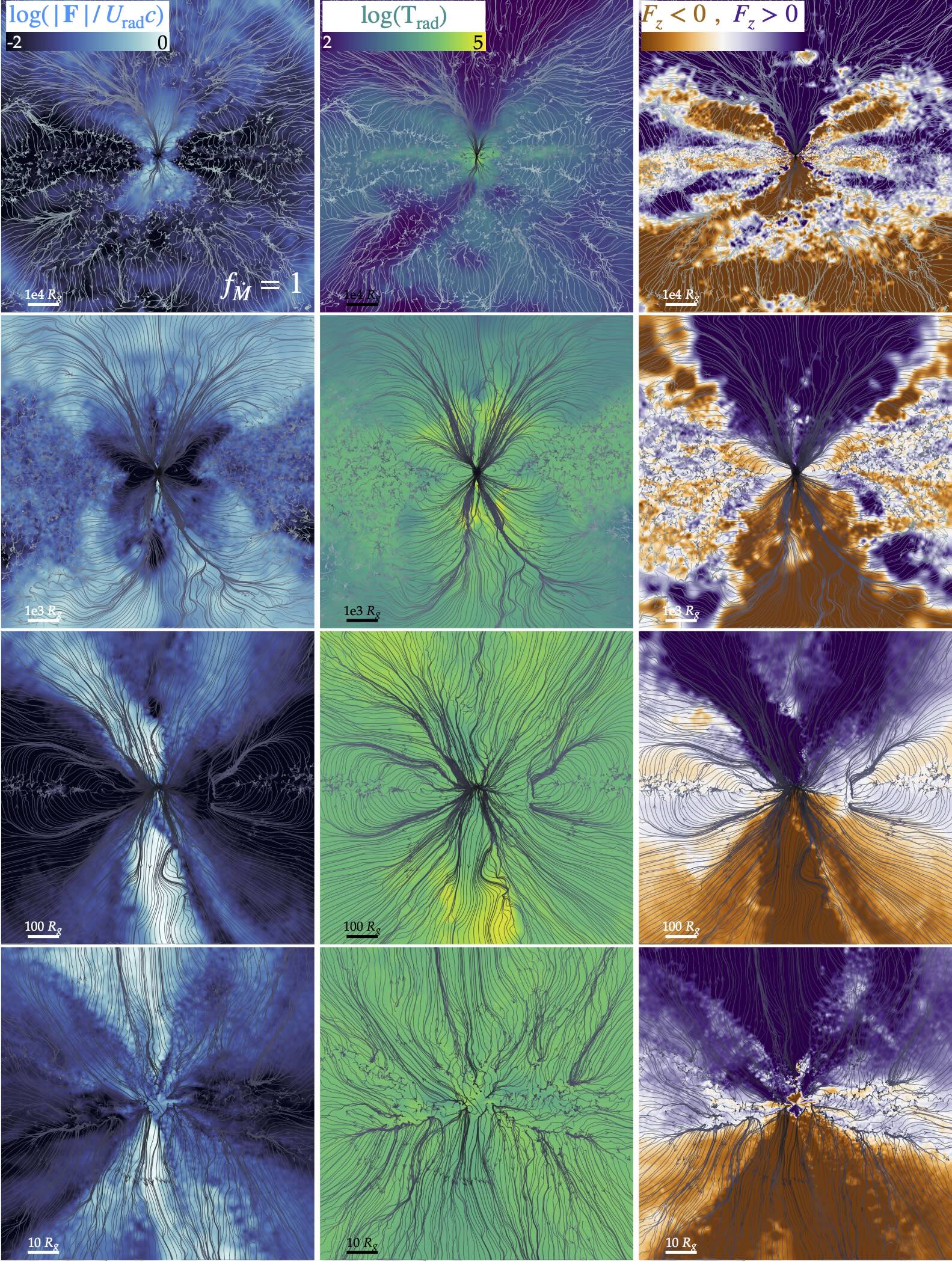}\hspace{5pt}\includegraphics[width=0.48\textwidth]{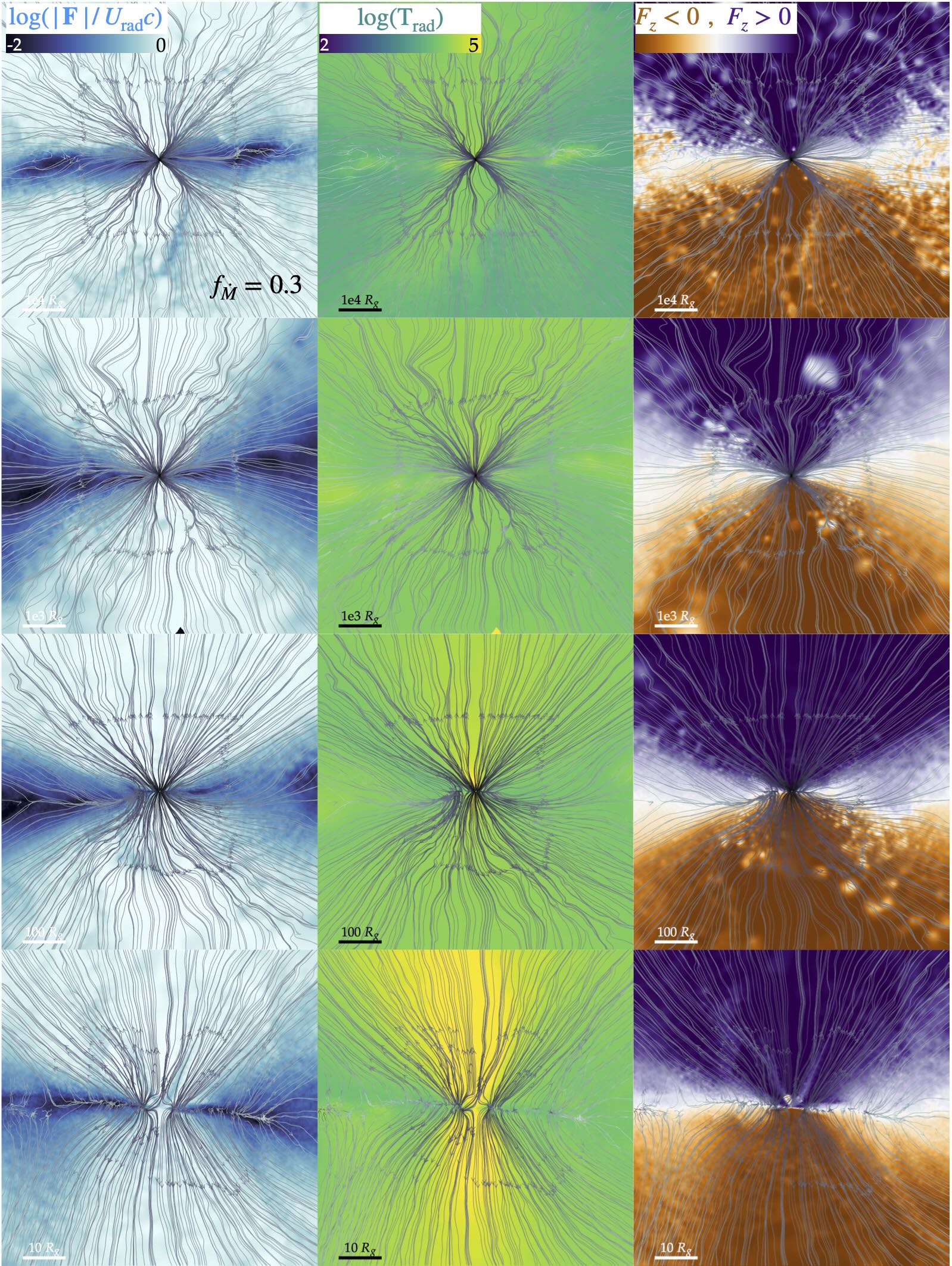}
	\caption{Edge-on radiation field ($x-z$ slice, BH in center) of the high-$\dot{m}$ simulations ($f_{\dot{M}}=1$ and $0.3$, labeled), at different scales, at the final time. Radiation flux ${\bf F}_{\rm rad}$ streamlines overlaid. Columns show reduced flux $|{\bf F}_{\rm rad}|/u_{\rm rad}c$ (free-streaming radiation at $\sim1$, trapped at $\ll 1$); radiation temperature $T_{\rm rad}$ and sign of vertical flux $F_{z}$. At $\dot{m} \gg 1$, the radiation is trapped in a thick, flared structure up to large heights, $T_{\rm rad}$ is cooler, and the outer disk is strongly illuminated by the inner (flux bends inwards to the disk via scattering in the polar regions). At $\dot{m} \sim 0.3$, the midplane traps radiation but it streams in the broad bicone, $T_{\rm rad}$ is slightly warmer with typical effective temperatures $\sim 3\times10^{4}$\,K in the polar direction, and there is no strong polar scattering bending flux back to the disk.\label{fig:maps.rad.streamlines}}
\end{figure*}

\section{Outgoing Radiation Properties \&\ Illumination}
\label{sec:rad}

\subsection{Radiative Efficiencies}
\label{sec:rad:eff}

\begin{figure*}
	\centering\includegraphics[width=0.49\textwidth]{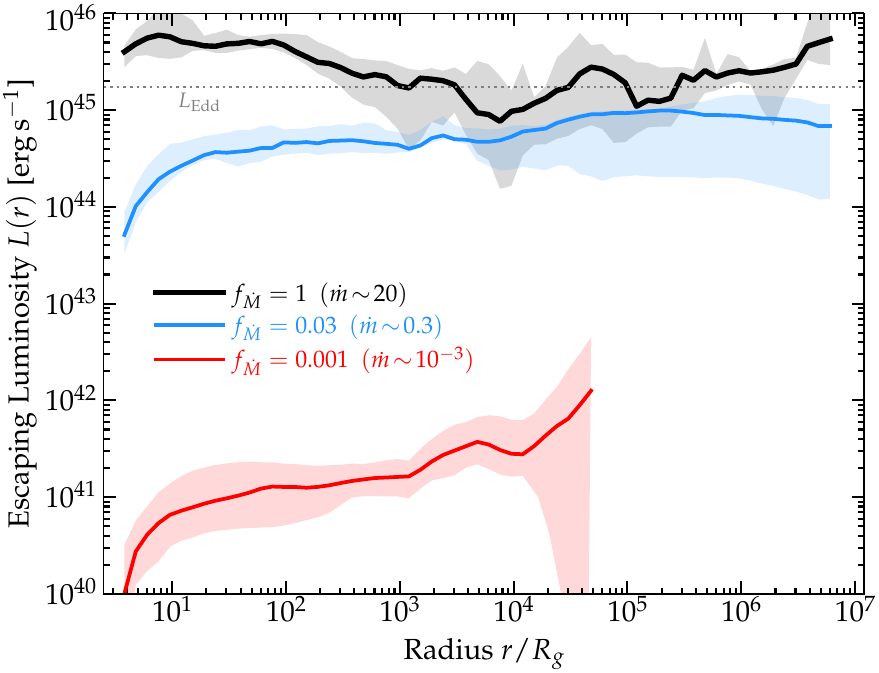}\includegraphics[width=0.49\textwidth]{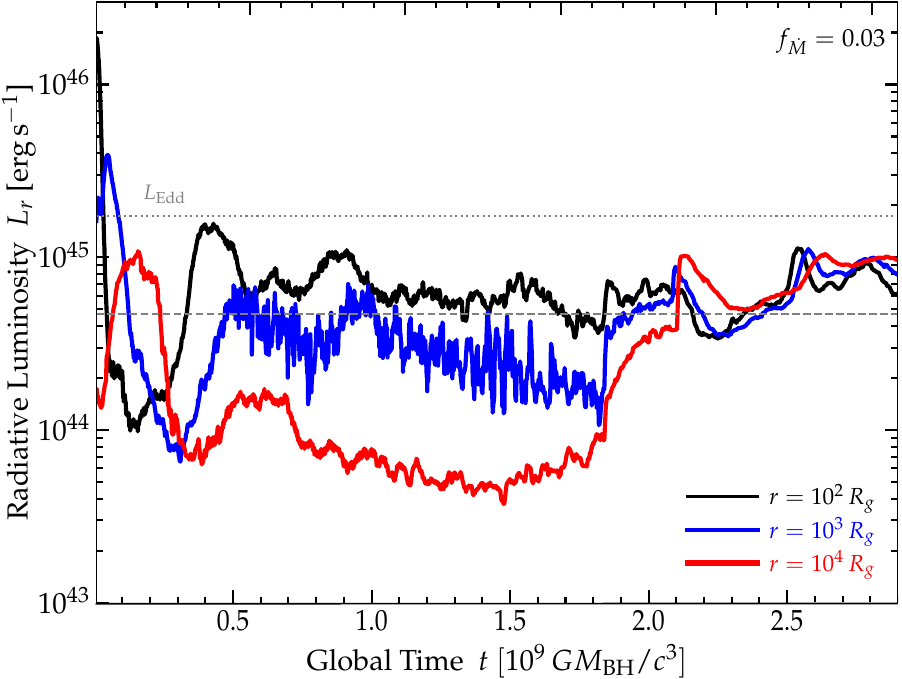}
	\caption{{\em Left:} Escaping (radial) radiative luminosity $L(r)$ vs.\ radius for the three runs (as Fig.~\ref{fig:mdot}; shaded shows inter-quartile range in time over last $\sim25\%$ of the runtime). We truncate the plot for the lowest-$\dot{m}$ where it appears to become dominated by ambient stellar radiation (we see comparable inward-and-outward radiation flux). The luminosities are relatively steady vs.\ $r$. {\em Right:} Luminosity vs.\ time for $\dot{m}\sim 0.3$. We see lags as changes from the inner disk propagate outwards. Dashed line shows a luminosity corresponding to $\epsilon_{r} \approx 0.1$ given the time-averaged $\dot{M}$ over this interval.\label{fig:lum}}
\end{figure*}

\begin{figure*}
	\centering\includegraphics[width=0.96\textwidth]{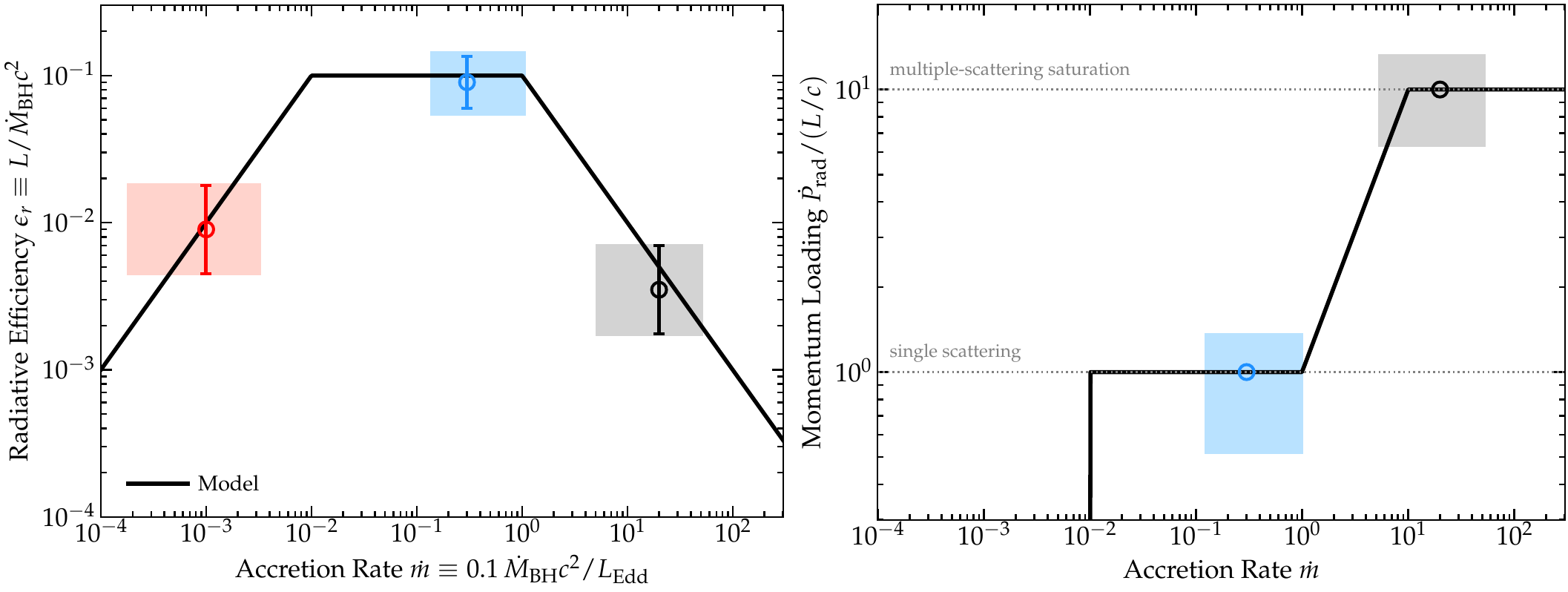}
	\caption{Comparison of simulations and the approximate subgrid model calibration in \S~\ref{sec:subgrid}, for radiative efficiency $\epsilon_{r}$ ({\em left}) and radiation-pressure driven wind momentum-loading ({\em right}). Points and boxes represent approximate mean values and ranges (sub-averaged in a few intervals $\sim10\%$ of the runtime) from our three simulations (per \S~\ref{sec:discussion:caveats}, these are intended for order-of-magnitude trends, not precise fits). The simulations are consistent with a model where $\epsilon_{r}$ peaks at $\sim 0.1$ for $0.01 \lesssim \dot{m} \lesssim 1$ and decreases $\propto\dot{m}$ at lower and $\propto\dot{m}^{-1}$ at higher $\dot{m}$, with no radiation pressure driven wind in the low-$\dot{m}$ (radiatively inefficient) state, winds following the approximate single-scattering expectation $\dot{P}_{\rm wind} \sim L/c$ at intermediate $\dot{m}$, increasing with some $\dot{P}_{\rm wind} \sim \langle \tau \rangle L/c \sim \dot{m} L/c$ up to a maximum of $\sim 10\,L/c$ or so in the highest-$\dot{m}$ phases we see.\label{fig:profiles.outflow.loadings}}
\end{figure*}

Consistent with \paperone\ and analytic expectations \citep{hopkins:superzoom.analytic,novikov.thorne:1973.astro.of.bhs,abramowicz:1988.slim.disks}, the radiative efficiency $\epsilon_{r} \equiv L / (\dot{M}_{\rm BH} c^{2})$ depends clearly but non-monotonically on $\dot{m}$. Here $L$ is the bolometric (summed over all RHD bands) net escaping luminosity, i.e.\ ${\bf F}_{\rm rad}\cdot\hat{\bf r}$ integrated over a sphere, evaluated at $\sim 100-1000\,R_{g}$, where essentially all the radiation has been generated but little has yet been reprocessed by the outer disk (Fig.~\ref{fig:lum}). The values quoted are time-averages, with factor-of-$\sim2$ scatter in time (Fig.~\ref{fig:profiles.outflow.loadings}), so the trends with $\dot{m}$ are robust but individual $\epsilon_{r}$ values should be taken as order-of-magnitude. For the fiducial, modestly sub-Eddington $\dot{m} \sim 0.1-0.4$ ($f_{\dot{M}} = 0.03$) case we find $\epsilon_{r} \sim 0.09$; for the super-critical $\dot{m} \sim 20-40$ ($f_{\dot{M}}=1$) case $\epsilon_{r} \sim 0.0035$, corresponding to $L \sim 1-2\,L_{\rm Edd}$, consistent with super-critical flows only exceeding $L_{\rm Edd}$ logarithmically as advection and outflows remove most of the accretion energy (see \paperone); and for the low-luminosity $\dot{m} \sim 0.001$ ($f_{\dot{M}}=0.001$) case $\epsilon_{r} \sim 0.009$.

The latter differs from \paperone, where the shorter evolution ($\sim 10^{5}\,t_{g}$, or $\sim 1\,\Omega^{-1}$ at $r \sim 10^{3}\,R_{g}$) did not allow the full $\dot{M} \propto r^{1/2}$ flow to develop, so $\dot{m}$ was $\sim 0.02-0.04$ (as opposed to $\sim 0.001$ here) with $\epsilon_{r} \sim 0.1-0.15$, similar to the $f_{\dot{M}} = 0.03$ case. The $\sim 10\times$ lower $\epsilon_{r}$ at $\dot{m} \sim 0.001$ here is consistent with $\epsilon_{r} \sim 0.1\,\dot{m}/\dot{m}_{\rm low}$ below $\dot{m}_{\rm low} \sim 0.01$, as expected from classic radiatively-inefficient, optically-thin, geometrically-thick accretion models \citep{narayan.yi.95:adaf.lowmass.bhs,narayan:bh.review.1998,yuan.narayan:2014.hot.accretion.review}.

\subsection{Emergent Spectral Energy Distributions}
\label{sec:rad:sed}

The emergent spectra also differ dramatically with $\dot{m}$. Over the radii where most of the radiation is produced, $\lesssim100-1000\,R_{g}$, at $\dot{m} \gtrsim 0.01$ (including the $\dot{m} \sim 0.02-0.04$ phase of $f_{\dot{M}} =0.001$ in \paperone, Fig.~16 therein) we find radiation temperatures $\sim (1-5) \times 10^{4}\,{\rm K}$, consistent with typical observed quasar spectra. Notably (as in \paperone), reprocessing means we do not see the naively-expected $T_{\rm rad} \propto (L/M_{\rm BH}^{2})^{1/4}$ scaling of razor-thin disk models (which is not observed; see \citealt{blackburne:2011.sizes.qso.acc.disks.microlensing.too.big,jimenez:2014.qso.disk.temp.profile.size.from.microlensing.large.flat,cai:2023.qso.sed.universal.w.flat.intrinsic.spectrum.into.uv.larger.than.expected.need.reprocessing.in.blr}); if anything (at fixed $M_{\rm BH}$), we see a weak inverse trend with $L$.

At $\dot{m} \sim 0.001$, by contrast, the inner-disk emission is primarily non-thermal hard X-rays ($T_{\rm rad} \gtrsim 10^{8}\,{\rm K} \sim 10\,{\rm keV}$). Our few-band treatment with a single evolved ``radiation temperature'' is not ideal for this: we treat absorption and re-emission as thermal, assigning re-emitted radiation $T_{\rm rad}^{\rm em} = T_{\rm gas}$ even where it should be non-thermal and much hotter, which almost certainly contributes to the sudden drop of $T_{\rm rad}$ outside the inner disk, and we miss Compton up/down-scattering and non-thermal radio/X-ray emission. Because the system is not extremely optically-thick, post-processing with more detailed radiative transfer is straightforward; but the qualitative statement that the radiation properties at $\dot{m} \ll 0.01$ are dramatically different from those at $\dot{m} \gg 0.01$ is likely robust, and consistent with analytic expectations and previous simulations of LLAGN \citep{narayan.yi.95:adaf.lowmass.bhs,yuan.narayan:2014.hot.accretion.review,guo:2022.superzoom.riaf.in.m87.nosf.nomhd.etc}. Our simulations recover this state change, and show that strong magnetization of the outer disk does not radically modify this intuition.

\subsection{Geometric \&\ Transient Effects}
\label{sec:rad:geom.time}

There is a clear geometric dependence of the escaping flux and spectrum (Fig.~\ref{fig:maps.rad.streamlines}). Most obviously, the outer disk is externally-illuminated by the inner disk, so at $R \gtrsim 1000\,R_{g}$, $T_{\rm rad}$ is highest in the polar regions (where inner-disk radiation escapes) and lowest in the midplane (where it equilibrates towards the gas or dust temperature in the atomic/molecular/dusty regions). At large radii where IR emission from the torus can escape (Bardati et al., in prep.) but the inner emission is shielded, $T_{\rm rad}$ is much lower, as it should be. There is also some hint of a ``receding torus'' at large-$L$.

The luminosities also vary clearly in time (Fig.~\ref{fig:lum}). For the reasons in \S~\ref{sec:sims:tradeoffs} we cannot make robust quantitative predictions for this, but we see it plainly even in the radial structure of the radiation flux: at any given time there are large changes at some radii, corresponding to radiation ``fronts'' from periods of higher/lower central $\dot{M}_{\rm BH}$, spread in time and space by reprocessing as they propagate outwards. More quantitative analysis requires post-processing without time-dilation.

\subsection{Self-Illumination and Thermo-Chemical Structure}
\label{sec:rad:illum}

Figs.~\ref{fig:maps.rad.streamlines} \&\ \ref{fig:profiles.rad} make it obvious that the outer disk is illuminated by the inner disk. At $\dot{m} \gtrsim 0.01$, within $|\cos{\theta}| \gtrsim 1/\sqrt{2}$ the radiation flux streamlines point outwards, but at lower latitudes they point {\em inwards} into the disk. Interior to $\sim 100\,R_{g}$, the streamlines instead rise from the midplane and extend outwards, as expected for the primary emission region. In the outer disk, the radiation flux points towards the midplane from above the ``top'' of the disk, out to $\gtrsim 10^{5}\,R_{g}$. This ``top'' lies at $|z| \sim R$, rather than the nominal density scale-height $|z|\sim H \sim (0.1-0.2)\,R$, simply because the disk lacks a sharp vertical ``edge'': per \paperone\ and \citet{hopkins:superzoom.disk}, the vertical density profile is a gradual power-law decline, akin to coronal/halo gas. The disk is illuminated even where it is not strongly flared (e.g.\ where $H/R$ is constant or slightly decreasing with $R$), because the flux lines ``bend'' towards the disk: the polar region is not entirely evacuated, but is better-ionized, and acts as a scattering layer which redirects a non-negligible fraction of the inner-disk flux onto the surface of the outer disk. At $\dot{m} \sim 0.001$, the outer disk is more obviously optically-thin (Fig.~\ref{fig:profiles.rad}), and the flux streamlines correspond to simpler radial outflow, so the disk is still illuminated but the illumination is less dramatically reprocessed.

The gas is generally not in LTE with the radiation, while in the outer disk dust and gas are well-coupled at temperatures below sublimation. To isolate the effect of illumination, we compare to the original ``zoom-in'' of \citet{hopkins:superzoom.overview} for $f_{\dot{M}}=1$, where the inner boundary was a simple sink at $\sim 300\,R_{g}$, so there was no inner disk to illuminate the outer disk (and the total luminosity was much lower). As expected, the most dramatic differences appear in the surface layers of the disk, where the gas temperature decreases towards the midplane, indicating that irradiation is an important heating source there. At higher $\dot{m}$ the midplane is well-shielded, so the effect on its thermochemistry is milder (as argued analytically in \citealt{hopkins:superzoom.analytic}), but not negligible: at $\sim 10^{3}-10^{4}\,R_{g}$, without illumination the mass-weighted gas temperature is a factor $\sim 10$ lower ($\sim 10^{3}\,{\rm K}$ vs.\ $10^{4}\,{\rm K}$).

The effect on the ionization of the disk is more dramatic: without illumination, the mass-averaged free electron fraction at $\sim 10^{3}-10^{5}\,R_{g}$ (sensitive to temperature and external UV/soft X-ray ionization) is a factor $\sim 10-100$ lower. As a result, the effective atomic/molecular transition and, importantly, the BLR move out by nearly a factor $\sim 10$ in $R$ with illumination, putting them in much better agreement with observations (as anticipated, since the BLR must be illuminated); the detailed properties of the predicted BLR will be compared to observations in future work (Bardati et al., in prep.). Also, as shown in Hopkins et al.\ (in prep.; see also \paperone), without illumination the original zoom-in has essentially no dust sublimation in the midplane. For dust grains subject to an un-shielded luminosity $L \sim L_{44}\,10^{44}\,{\rm erg\,s^{-1}}$ (flux $\sim L/(4\pi r^{2})$) with comparable absorption/emission efficiencies $Q_{\rm abs} \sim Q_{\rm em}$ and sublimation temperature $T_{\rm sub} \sim 2000\,{\rm K}$, the usual estimate of the sublimation radius is $r_{\rm sub} \sim (L/16\pi\sigma_{B} T_{\rm sub}^{4})^{1/2} \sim 0.01\,L_{44}^{1/2}\,{\rm pc} \sim 10^{4} L_{44}^{1/2} R_{g}$, which is reasonably similar to what we see in our full simulations, at least for the dust above the midplane. In the midplane, the incident flux being smaller means that dust persists further in, so there is not a simple very sharp cutoff in the mass-weighted dust-to-gas-ratio versus $R$, but a more gradual decrease (as a deeper surface layer is sublimated) until about an order-of-magnitude smaller radius, where the thermal temperature of the midplane is sufficiently hot (and the dust-gas collisional coupling sufficiently strong) to sublimate the dust, very similar to the prediction in the analytic models in \citet{hopkins:superzoom.analytic}.
All of these thermochemical results also give further confidence to the arguments in \citet{hopkins:superzoom.disk,hopkins:superzoom.analytic} that non-ideal MHD effects (ambipolar drift, Hall current, Ohmic resistivity) are negligible on the scales here (sub-dominant by $>10$ orders of magnitude).

\begin{figure}
	\centering\includegraphics[width=0.99\columnwidth]{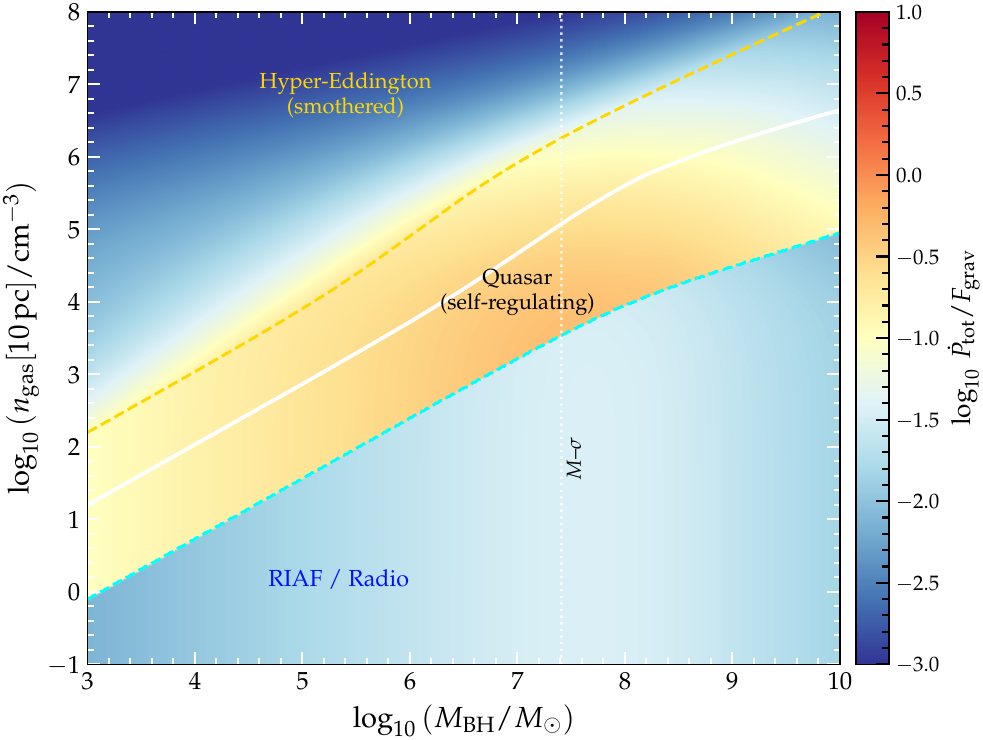}
	\caption{Illustrative AGN feedback regime diagram in the space of BH mass vs.\ nuclear gas density $n_{\rm gas}(10\,{\rm pc})$ for $\sigma_{\ast} = 200\,{\rm km\,s^{-1}}$, using the sub-grid scalings of \S~\ref{sec:subgrid} (calibrated on the three simulations here; the broad trends, not the precise contour locations, are the robust content). Colors show $\dot{P}_{\rm tot}/F_{\rm grav}$, the ratio of total feedback momentum flux to gravity. Contours mark $\dot{m} = 0.01$ ({\em cyan}, RIAF/radio transition), $\dot{m} = 1$ ({\em white}, Eddington), and $\dot{m} = 10$ ({\em gold}, hyper-Eddington onset). At low density, accretion is radiatively inefficient and jets dominate (RIAF/radio mode); at intermediate densities radiation-pressure feedback self-regulates ($\dot{P}\sim F_{\rm grav}$, quasar mode); at very high densities the flow is hyper-Eddington and feedback is ``smothered.'' The $M_{\rm BH}$--$\sigma$ mass is marked (vertical line).\label{fig:subgrid.regimes}}
\end{figure}

\section{Towards a Sub-Grid Prescription for High and Low-State Accretion}
\label{sec:subgrid}

Per \S~\ref{sec:intro}, state-of-the-art high-resolution simulations of galaxy/ISM/cosmological scales barely resolve the BH ROI ($\sim 1-10\,$pc for supermassive BHs), and many do not even resolve this scale, invoking prescriptions such as Bondi-Hoyle accretion or more general models for BH accretion in a non-Keplerian potential allowing for gravitational torques by e.g.\ bars and spiral arms  \citep{hopkins:inflow.analytics,fgq2012,angles.alcazar:grav.torque.accretion.cosmo.sim.implications}, that are designed to estimate the accretion rate into the ROI ($\dot{M}_{\rm in}^{\rm ROI}$) from large-scale properties.\footnote{Like most galaxy simulations, we work in terms of the ROI, $R_{\rm ROI} \sim G M_{\rm BH}/\sigma_{\ast}^{2}$, not the Bondi radius $R_{B} \sim G M_{\rm BH}/c_{\infty}^{2}$. In most high-$\dot{m}$ systems, accretion is cold ($c_{s} \ll \sigma_{\ast}$ for the accreting gas) and limited by angular momentum, not thermal pressure \citep{hopkins:inflow.analytics,hopkins:superzoom.overview}, and the ISM in galactic nuclei (even in low-$\dot{m}$ systems like M87$^{\ast}$ or Sgr A$^{\ast}$, which may be in a Bondi-like state) is multi-phase, with factor $\sim 10^{4}$ variation in $R_{B}$ depending on ``which gas'' one considers.} With that in mind, we can use the insights from the simulations here to propose some qualitative scalings for said models. In particular, while the accretion and jet scalings for $\dot{m} \ll 1$ from horizon-to-BHROI have been studied more extensively already by \citet{yuan.narayan:2014.hot.accretion.review} and \citet{cho:2023.multiscale.accretion.sims.bondi.inflow.model,guo:2024.fluxfrozen.disks.lowmdot.ellipticals}, there has not been self-consistent work thus far on the high-$\dot{m}$, strongly-magnetized regime reaching large scales with the necessary physics of multi-phase, NLTE multi-group optically thick+thin RHD.

\subsection{Basic Scalings}
\label{sec:subgrid:scalings}

Given some $\dot{M}_{\rm in}^{\rm ROI} = \dot{M}_{\rm in}(r \sim R_{\rm ROI})$ defined at the ROI, the accretion rate onto the BH can be estimated from \S~\ref{sec:steady.state:acc} as: 
\begin{align}
\dot{M}_{\rm BH} \sim &
\begin{cases}
 \left( \frac{R_{g}}{R_{\rm ROI}} \right)^{0.1} \dot{M}_{\rm in}^{\rm ROI} & (\dot{M}_{\rm in}^{\rm ROI} \ge \dot{M}_{c}) \\ 
 \left( \frac{R_{g}}{R_{\rm ROI}} \right)^{1/2} \dot{M}_{\rm in}^{\rm ROI}  &  (\dot{M}_{\rm in}^{\rm ROI} < \dot{M}_{c}) \\
\end{cases} 
\end{align}
where $\dot{M}_{c} \sim 0.01\,\dot{M}_{\rm crit}$ (corresponding to $\dot{m} \sim 0.01$, see e.g. \citealt{Cho2022}). As discussed therein the $1/2$ scaling at low-$\dot{m}$ (in the hot accretion flow regime) is well studied, here we augment this with the weaker scaling for the disky (strongly-magnetized) regime. The time-averaged total outflow mass coming ``back out'' at the ROI must be: 
\begin{align}
\dot{M}_{\rm wind} = \dot{M}_{\rm out}^{\rm ROI} &\sim \dot{M}_{\rm in}^{\rm ROI} - \dot{M}_{\rm BH} \ .
\end{align}

Given $\dot{M}_{\rm BH}$, the key feedback inputs can be parameterized in terms of ``efficiency factors.'' The radiative efficiency $\epsilon_{r}$ can be approximated by: 
\begin{align}
\epsilon_{r} \equiv \frac{L}{\dot{M}_{\rm BH} c^{2}} \sim & 
\begin{cases}
 0.1\,\dot{m}^{-1} & (\dot{m} \ge 1) \\ 
 0.1 & (\dot{m}_{\rm low} < \dot{m} < 1) \\ 
 0.1\,(\dot{m}/\dot{m}_{\rm low}) & (\dot{m} \le \dot{m}_{\rm low}) \\
\end{cases}
\end{align}
where we recall our definition $\dot{m} \equiv 0.1\,\dot{M}_{\rm BH} c^{2} / L_{\rm Edd}$, and the sub-critical boundary is approximately $\dot{m}_{\rm low} \sim 0.01$. This is qualitatively consistent with standard scalings for the optically-thin hot-flow regime at low-$\dot{m}$ \citep{narayan.yi.95:adaf.lowmass.bhs,yuan.narayan:2014.hot.accretion.review}, canonical efficiencies for radiative disks at intermediate $\dot{m}$, and inefficient accretion again owing to radiative trapping effects at large $\dot{m}$ (effectively ensuring that $L \sim L_{\rm Edd}$ even at $\dot{m} \gtrsim 1$; \citealt{abramowicz:1988.slim.disks,jiang:2019.superedd.sims.smbh.prad.pmag.modest.outflows,jiang.dai:2024.supereddington.accretion.simulations.review}). 

The kinetic wind component -- which we distinguish from a Blandford-Znajek type jet -- is conveniently parameterized in the literature in terms of its momentum-loading $\dot{P}_{\rm wind}$, and for the reasons in \S~\ref{sec:outflow} we decompose it into a radiatively-driven outflow $\dot{P}_{\rm rad,\,w}$ and ``turbulent'' outflow $\dot{P}_{\rm turb,\,w}$, 
\begin{align}
\dot{P}_{\rm wind} &= \dot{P}_{\rm out}^{\rm ROI} = \dot{P}_{\rm rad,\,w} + \dot{P}_{\rm turb,\,w} \ .
\end{align}
The radiation-pressure driven outflow scales as: 
\begin{align}
\dot{P}_{\rm rad,\,w} \sim & 
\begin{cases}
\dot{m}_{\rm sat} (L/c) & (\dot{m}_{\rm sat} \le \dot{m}) \\ 
\dot{m} (L/c) & (1 \le \dot{m} < \dot{m}_{\rm sat}) \\ 
 L/c & (\dot{m}_{\rm low} < \dot{m} < 1) \\ 
0 & (\dot{m} \le \dot{m}_{\rm low}) \\
\end{cases} 
\end{align}
with $\dot{m}_{\rm sat} \sim 10$. This corresponds to an average mass-weighted wind velocity: 
\begin{align}
\langle v^{\rm rad}_{\rm wind} \rangle \sim & 
\begin{cases}
0.03\,c\,m_{7}^{0.05}\,(\dot{m}_{\rm sat}/\dot{m}) & (\dot{m}_{\rm sat} \le \dot{m}) \\ 
0.03\,c\, m_{7}^{0.05} & (\dot{m}_{\rm low}  < \dot{m} < \dot{m}_{\rm sat}) \\ 
- & (\dot{m} \le \dot{m}_{\rm low}) \ . \\
\end{cases} 
\end{align}
Essentially, this gives negligible radiation-pressure-driven outflow escaping past $R_{\rm ROI}$ in the radiatively-inefficient low-$\dot{m}$ state; the standard $\sim L/c$ expectation for radiation-pressure driven outflows where an order-unity fraction of the radiation is captured but multiple-scattering of the broad-wavelength continuum is not critical; and an increase in the super-critical regime (where the flow {\em does} become optically thick to its own continuum over most of the SED) as $\dot{m} L/c \sim \langle \tau \rangle L/c$, up to a saturation level $\sim 10\,L/c$ (reflecting the well-known tendency of multiple-scattering to self-saturate; see \citealt{thompson:rad.pressure,krumholz:2012.rad.pressure.rt.instab,davis:2014.rad.pressure.outflows,tsang:monte.carlo.rhd.dusty.wind,hopkins:2019.grudic.photon.momentum.rad.pressure.coupling}).

The ``turbulent'' outflow has mass entrained and decelerated at all radii, mixing as in a standard $\rho \propto r^{-1}$-law, spherical accretion flow (which conserves momentum flux) requiring 
\begin{align}
\dot{P}_{\rm turb,\,w} &\sim \dot{M}_{\rm wind} v_{\rm esc}(R_{\rm ROI}) \\
\nonumber & \approx \dot{M}_{\rm wind} \sqrt{\frac{8\, G M_{\rm BH}}{R_{\rm ROI}}}
\sim 3 \dot{M}_{\rm wind} \sigma_{\ast}^{\rm ROI}\ .
\end{align}
This will dominate the momentum flux in the low state, or extremely supercritical states $\dot{m} \gtrsim 300 m_{7}^{0.05}/\sigma_{100}$ (with $\sigma_{100} \equiv \sigma_{\ast}^{\rm ROI}/100 {\rm km\,s^{-1}}$; or $\dot{M}_{\rm in}^{\rm ROI} \gtrsim 300\,{\rm M_{\odot}\,yr^{-1}}\,m_{7}/\sigma_{100}$) where the radiation-pressure driven outflow from smaller scales is smothered by the columns at large-$R$. The latter is likely only relevant for phenomena such as hyper-Eddington accretion onto IMBH seeds, but this could be important where such phenomena has been proposed to account for e.g.\ the properties of AGN such as ``little red dots'' (LRDs) observed at high redshifts \citep{matthee:2024.little.red.dots,kokorev:2024.little.red.dots}. Indeed for the scenario proposed in \citet{shi:2024.seed.to.smbh.case.study.subcluster.merging.pairing.fluxfrozen.disk} for LRDs via strongly-magnetized accretion onto a $\sim 10^{6}\,{\rm M}_{\odot}$ BH in the center of a dense natal star cluster/proto-bulge, the smothered regime would require only $\dot{M}_{\rm in}^{\rm ROI} \gtrsim 0.1\,{\rm M_{\odot}\,yr^{-1}}$, easily satisfied in those simulations. 

Since we do not follow jets but wish to compare them for completeness, we can consider a standard prescription. The energy is determined by the magnetization near-horizon, following the standard \citet{blandfordznajek:1977} scaling as in \citet{tchekhovskoy:2011.mad.disk.jets} with power $\propto (a\,\Phi_{B}/M)^{2}$ or, in terms of efficiency (using $\dot{M} \sim 2\pi \Sigma_{\rm gas} R v_{\rm K} \sim  2\pi \Sigma_{\rm gas} R c$ within the ISCO)
\begin{align}
\frac{\dot{E}_{j}}{\dot{M}_{\rm BH} c^{2}} \approx \frac{\dot{P}_{j}}{\dot{M}_{\rm BH} c} \sim \left( \frac{a B}{B_{\rm crit}}\right)^{2}  \sim \left( \frac{a v_{A}}{v_{\rm K}}\right)^{2} \sim \left( \frac{a H}{R}\right)^{2}_{R_{\rm ISCO}}
\end{align}

The kinetic efficiency $\epsilon_{\rm k} \equiv (1/2)\dot{M}_{\rm wind} \langle v_{\rm wind} \rangle^{2} / \dot{M}_{\rm BH} c^{2}$ is therefore
\begin{align}
\epsilon_{\rm k} \sim & 
\begin{cases}
1.7 \times 10^{-6} \sigma_{100}^{9/5} & (290\,\sigma_{100}^{-4/5} \le \dot{m}) \\ 
0.0014 \sigma_{100}^{1/5}\,(\dot{m}_{\rm sat}/\dot{m})^{2} & (\dot{m}_{\rm sat} \le \dot{m} < 290\,\sigma_{100}^{-4/5}) \\ 
0.0014 \sigma_{100}^{1/5} & (\dot{m}_{\rm low} < \dot{m} \le  \dot{m}_{\rm sat} ) \\ 
0.0011 \sigma_{100} & (\dot{m} \le \dot{m}_{\rm low} ) \ . \\ 
\end{cases} 
\end{align}
The thermal and magnetic-loading are generally sub-dominant or at most comparable so we simply parameterize them as: 
$\epsilon_{\rm th} \equiv \dot{M}_{\rm wind} \langle u_{\rm wind}^{\rm thermal} \rangle / \dot{M}_{\rm BH} c^{2}$ and 
$\epsilon_{\rm B} \equiv \dot{M}_{\rm wind} \langle u_{\rm wind}^{\rm B} \rangle / \dot{M}_{\rm BH} c^{2}$: 
\begin{align}
\epsilon_{\rm th} \sim & 
\begin{cases}
0.1\,\epsilon_{\rm k} & (\dot{m}_{\rm low} < \dot{m} )  \\ 
\epsilon_{\rm k} & (\dot{m} \le \dot{m}_{\rm low} )  \\ 
\end{cases} 
\end{align}
and
\begin{align}
\epsilon_{\rm B} \sim & 
\begin{cases}
0.25\,\epsilon_{\rm k} & (\dot{m}_{\rm low} < \dot{m} )  \\ 
0.03\,\epsilon_{\rm k} & (\dot{m} \le \dot{m}_{\rm low} )  \\ 
\end{cases} 
\end{align}
(a very crude average over our runs at large-$r$). 
For low accretion rates, this result is consistent with the GRMHD simulations of \citet{Cho2026}, which showed that $\epsilon_{\rm k} \sim \epsilon_{\rm th} \gg \epsilon_{\rm B}$ measured at BHROI across BH spins $a=0-0.9$.

\subsection{Implications at Typical ISM/Galactic Scales}
\label{sec:subgrid:example}

To illustrate the impact of these scalings, consider a common ``high'' resolution for cosmological galaxy formation/ISM simulations, $\Delta x \sim 10\,$pc, with an isothermal-like central potential with $\sigma_{\ast} \sim 200\,{\rm km\,s^{-1}}$ (from stars+dark matter), some gas density $n_{\rm gas}$ measured within the central $R_{\rm res} \sim \Delta x \sim 10\,{\rm pc}$, and a BH of mass $M_{\rm BH}$ at the center. Since $R_{\rm res} > R_{\rm ROI}$ for smaller SMBH masses, we use the gravitational-torque sub-grid scalings calibrated on resolved simulations of accretion over the intermediate, non-Keplerian scales $\sim 0.01-10\,$pc \citep{daa:20.hyperrefinement.bh.growth}: $\dot{M}_{\rm in}^{\rm ROI} = \eta_{\rm g} f_{\rm fb} M_{\rm gas}\Omega(R)$, where $\eta_{\rm g} \sim  0.01\,(M_{\rm BH}/M_{d})^{1/6} / [1+3 (M_{d}/10^{9} {\rm M}_{\odot})^{1/3} M_{\rm gas}/M_{d}]$, $M_{d} \sim M_{\rm gas} + f_{d} M_{\ast}$ with $M_{\ast} \sim \sigma_{\ast}^{2} R/G$ and typical (from those simulations) assumed $f_{d} \sim 0.05$ being a measure of rotational support, and $f_{\rm fb} \equiv \Sigma_{\rm eff}/(\Sigma_{\rm crit} + \Sigma_{\rm eff})$ with $\Sigma_{\rm eff} = M_{\rm enc}/\pi R^{2}$ and $\Sigma_{\rm crit} \sim 3000\,{\rm M_{\odot}\,pc^{-2}}$ reflecting the efficacy of ejection by stellar feedback (with $M_{\rm enc} = M_{\ast}(R) + M_{\rm gas} + M_{\rm BH}$). With that sub-grid model for accretion from $R \sim \Delta x$ to the ROI, we then apply the scalings above to calculate $\dot{m}$, $L$, $\dot{P}_{\rm wind}$, etc. It is useful to compare the emergent feedback to that needed to eject the gas at this larger scale. In momentum terms, since $\dot{P}$ is just a force, we compare it to the gravitational force binding the gas at $R \sim \Delta x$: $\dot{P}_{\rm crit} = F_{g}(R \sim \Delta x) \sim G M_{\rm enc} M_{\rm gas} / R^{2}$. 

We see three qualitatively distinct regimes, illustrated in Fig.~\ref{fig:subgrid.regimes}:
\begin{enumerate}

\item Below a critical $n_{\rm low} \sim n_{\rm gas}(10\,{\rm pc}) \sim 1300\,m_{7}^{5/6}\Delta x_{10}^{-11/6} \sigma_{300}^{-2/3}\,{\rm cm^{-3}}$, $\dot{M}_{\rm ROI} \lesssim \dot{M}_{c}$ and accretion becomes much less efficient, so $\dot{m}$ drops rapidly while (owing to decreasing $\epsilon_{r}$) $L$ drops even more rapidly. The local/self-regulating or ``quasar-mode'' feedback in this low-$\dot{m}$ regime is weak, $\dot{P} \ll \dot{P}_{\rm crit}$, but the BH will not grow much in this state since $\dot{m} \ll 0.01$ implies mass-doubling times longer than the Hubble time. Energetic feedback, on the other hand, could be very important, as the kinetic efficiency is relatively high and jet energy could easily escape the central regions per \S~\ref{sec:outflow:low.mdot:jets} and heat larger scales -- the ``radio'' or ``maintenance'' mode of AGN feedback \citep{croton:sam,ciotti:recycling.with.feedback.rad.vs.mech,weinberger:2017.agn.fb.tng}. 

Interestingly, in this regime the jet power scales as $\sim a^{2} 10^{39.5}\,{\rm erg\,s^{-1}}\,n_{1} \sigma_{300}^{5/3} R_{10}^{11/6} (m_{7}/f_{d,\,0.05})^{1/6}$, or $\sim a^{2} 10^{45.5}\,{\rm erg\,s^{-1}}\,f^{5/6}_{\rm gas}(\Delta x)\,(m_{7}/\Delta x_{10})^{1/6}\, \sigma_{300}^{11/3}$ -- i.e.\ because of the scalings of $\dot{M}$, etc., it depends strongly on $n$ and $\sigma_{\ast}$, as $\propto \sigma_{\ast}^{4}$ roughly, but weakly on $M_{\rm BH}$ directly, suggesting that sustaining powerful jets in deep potential wells of massive ellipticals is relatively ``easy.''

\item Between $n_{\rm low} \lesssim n_{\rm gas} \lesssim n_{\rm high} \sim 2.4 \times 10^{5}\,{\rm cm^{-3}}\,f_{d,\,0.05}^{1/6}\,m_{7}^{5/6}\,\Delta x_{10}^{-11/6}\,\sigma_{300}^{-13/15}$, accretion is in the ``typical'' AGN regime with $0.01 \lesssim \dot{m} \lesssim 10$. In this state, BH growth is rapid but $\dot{P} \sim \dot{P}_{\rm crit}$, so winds tend to blow out gas without allowing runaway BH growth. This is the canonical ``quasar mode'' or ``self-regulating'' regime.

\item At $n \gg n_{\rm high}$, the system enters a ``hyper-Eddington'' mode with $\dot{m} \gg 10$, where feedback becomes saturated/smothered. The luminosity caps near Eddington ($L \sim L_{\rm Edd}$), and radiation-driven outflows saturate at their maximum efficiency ($\dot{P} \sim \dot{m}_{\rm sat}\,L/c$). Strong outflows exist in the inner disk but cannot break out through the massive outer envelope; outer-scale outflows are more turbulent than coherently driven. This resembles a ``quasi-star'' configuration \citep{shi:2022.hyper.eddington.no.bhfb}. The BH growth timescale in this regime is very short ($\lesssim$\,Myr), so the BH mass increases rapidly until it enters the ``self-regulating'' band.

\end{enumerate}

In practice, the trajectory through these regimes is complex. But as a schematic, one can imagine a BH of small seed mass in a dense star cluster or proto-bulge. At early times with high nuclear gas density, the BH accretes in the hyper-Eddington regime, growing rapidly at approximately fixed $n(\Delta x)$ until $M_{\rm BH}$ becomes large enough that feedback can regulate the gas supply (entering the self-regulating band). The BH mass at which this transition occurs corresponds approximately to the $M_{\rm BH}-\sigma$ scaling predicted by momentum-driven feedback arguments \citep{silkrees:msigma,king:msigma.superfb.1,murray:momentum.winds,hopkins:bhfp.theory}: for $n_{\rm gas} \sim 10^{3},\,10^{5},\,10^{7}\,{\rm cm^{-3}}$ we estimate self-regulation kicks in at $M_{\rm BH} \sim 10^{4},\,10^{6},\,10^{8}\,{\rm M_{\odot}}$, respectively. After blowout, gas densities drop below $n_{\rm low}$ and the system transitions to the low-$\dot{m}$ state, where jet/radio-mode feedback may dominate.

Note that in basically all cases $\dot{M}_{\rm out} \sim \dot{M}_{\rm in}^{\rm ROI}$, but the suppression of the accretion rate onto the BH differs enormously between regimes: at $\dot{m} \gtrsim 0.01$, $\dot{M}_{\rm BH} \sim (R_{g}/R_{\rm ROI})^{0.1}\,\dot{M}_{\rm in}^{\rm ROI} \sim (0.1-0.2)\,\dot{M}_{\rm in}^{\rm ROI}$, while at $\dot{m} \lesssim 0.01$, $\dot{M}_{\rm BH} \sim (R_{g} / R_{\rm ROI})^{1/2} \dot{M}_{\rm in}^{\rm ROI} \sim 10^{-3}\,\dot{M}_{\rm in}^{\rm ROI}$. Since $\epsilon_{r}$ also decreases $\propto \dot{m}$ at $\dot{m} \lesssim 0.01$, the luminosity decreases by a ``double factor'': while $L$ decreases by only a factor of a few from $f_{\dot{M}}=1$ ($\dot{m} \sim 10-40$) to $f_{\dot{M}}=0.03$ ($\dot{m} \sim 0.1-0.4$), it decreases by another factor of $\sim 1000$ to $f_{\dot{M}}=0.001$ ($\dot{m} \sim 0.001$).

\section{Discussion}
\label{sec:discussion}

\subsection{Caveats and Limitations}
\label{sec:discussion:caveats}

{\em Time dilation:} As discussed in \S~\ref{sec:sims:tradeoffs}, the time-dilation method sacrifices information about variability on timescales shorter than $a_{i}^{-1}\,\Omega^{-1}$ at each radius. So while quasi-steady-state properties (time-averaged profiles, outflow rates, etc.) should be reliable, predictions for variability, flaring, or transients on timescales comparable to the local dynamical time at small radii should be treated with caution. Our tests with different dilation parameters, and comparison with the un-dilated \paperone\ simulations, give some confidence that the basic quasi-steady state is not an artifact of the method.

{\em Lack of full GR and jet physics:} Our Newtonian gravity with a Paczy{\'n}ski-Wiita potential and simple sink inner boundary misses the proper pileup of poloidal magnetic flux at the horizon \citep[see][]{kaaz:2024.hamr.forged.fire.zoom.to.grmhd.magnetized.disks}, frame-dragging, and a self-consistent Blandford-Znajek jet, and our sub-grid jet injection is a crude approximation. This particularly affects the low-$\dot{m}$ MAD state at $\lesssim 10-100\,R_{g}$ and jet properties; a full GRMHD treatment joined to our multi-physics simulation at some intermediate radius (see e.g.\ \citealt{kaaz:2024.hamr.forged.fire.zoom.to.grmhd.magnetized.disks}) would be the ideal next step.

{\em Radiation treatment:} Our few-band M1 RHD with an evolved $T_{\rm rad}$ does not capture resonant line scattering (which may be important for UV line-driven winds; \citealt{murray:1995.acc.disk.rad.winds,proga:disk.winds}), nor detailed non-thermal emission relevant for the hard X-ray corona and radio emission at low $\dot{m}$ (\S~\ref{sec:rad:sed}). Post-processing with Monte Carlo radiative transfer codes could help quantify these uncertainties.

{\em Resolution:} While exceptional by the standards of cosmological and galaxy-scale simulations, our resolution remains far coarser than dedicated shearing-box and GRMHD simulations. Convergence with respect to the resolved scale-height is always a concern; \paperone\ found basic disk properties reasonably well-converged at our fiducial resolution, but finer-scale turbulent structure may still be under-resolved in some regimes.

{\em We study one example:} Our simulations all use a single BH mass $M_{\rm BH} \sim 1.3 \times 10^{7}\,{\rm M_{\odot}}$ at a specific epoch in a specific galaxy. Crucially, this means the sub-grid scalings of \S~\ref{sec:subgrid} -- the efficiencies $\epsilon_{r}$, $\epsilon_{k}$, the loadings $\dot{P}_{\rm wind}$, and the regime diagram (Fig.~\ref{fig:subgrid.regimes}) -- are anchored by effectively {\em three} measured points (the three $f_{\dot{M}}$ runs) at one mass and epoch, with the mass-, $\sigma_{\ast}$-, and $\dot{m}$-dependences taken from analytic/previous-simulation expectations rather than directly mapped here. The diagram and piecewise formulae should therefore be read as encoding the qualitative {\em regimes} and order-of-magnitude trends, not as precisely-calibrated fits; the locations of the regime boundaries are uncertain by factors of several. Extrapolation to other masses, redshifts, and environments should be done cautiously, and would benefit greatly from a grid of simulations spanning $M_{\rm BH}$ and environment in future work.

\subsection{Comparison with Previous Work}
\label{sec:discussion:comparison}

Our results for the high-$\dot{m}$ magnetically-dominated disk are consistent with and extend the findings of \paperone\ and \citet{hopkins:superzoom.disk,hopkins:superzoom.analytic}. The key advance here is the demonstration that these disks survive in quasi-steady-state over timescales $\gtrsim 10^{9}\,t_{g}$, long enough for outflows and radiation to propagate from the inner disk to the ROI and beyond, and the quantification of the resulting feedback properties.

For the low-$\dot{m}$ regime, the transition to a hot, geometrically-thick, MAD-like inner flow surrounded by a strongly-magnetized outer disk is qualitatively very similar to \citet{guo:2024.fluxfrozen.disks.lowmdot.ellipticals} (see also \citealt{guo:2022.superzoom.riaf.in.m87.nosf.nomhd.etc}), who simulated gas accreting from hot halos in massive ellipticals; our simulations add the multi-phase thermo-chemistry, star formation, and radiation required for the high-$\dot{m}$ regime.

The radiative efficiencies and outflow properties we find are broadly consistent with classic analytic expectations \citep{shakurasunyaev73,narayan.yi.95:adaf.lowmass.bhs,abramowicz:1988.slim.disks}, but their detailed properties -- particularly the spatial distribution of outflow launching, the role of opacity transitions, and the self-consistent illumination of the outer disk -- go beyond what simple analytic models predict.

\subsection{Implications for Observations}
\label{sec:discussion:obs}

Several of our predictions are testable with current and upcoming observations. (1) The strong illumination of the outer disk by the inner accretion flow is consistent with recent spatially-resolved observations of BLR structure by the GRAVITY collaboration \citep{gravity:2018.sturm.blr.rotating.thick.disk,gravity:2020.resolved.blr.size.disk.inside.dust.sub,gravity:2021.resolved.blr.disk.hot.dust.coronal.regions}, and the prediction that the BLR is pushed outward by illumination could be tested by systematic comparisons of reverberation mapping radii with self-consistent disk models. (2) The radiation-driven outflow velocities ($\sim 5000-30,000\,{\rm km\,s^{-1}}$) and momentum-loading ($\sim L/c$ at sub-Eddington and $\sim 10\,L/c$ at super-Eddington) are consistent with observations of broad absorption line quasars and ultra-fast outflows \citep{dunn:agn.fb.from.strong.outflows,faucher-giguere:2012.felobal.model,sturm:2011.ulirg.herschel.outflows}. (3) The transition from radiation-dominated to jet-dominated feedback at $\dot{m} \sim 0.01$ is consistent with the observed dichotomy between radiatively-efficient ``quasar mode'' and radiatively-inefficient ``radio mode'' AGN \citep{yuan.narayan:2014.hot.accretion.review}. (4) The smothered hyper-Eddington regime, where the emergent luminosity caps near $L_{\rm Edd}$ despite $\dot{m} \gg 1$, may be relevant for understanding ``little red dots'' and other high-redshift AGN with unusual spectral properties \citep{matthee:2024.little.red.dots,kokorev:2024.little.red.dots,shi:2024.seed.to.smbh.case.study.subcluster.merging.pairing.fluxfrozen.disk}.

\section{Conclusions}
\label{sec:conclusions}

We have extended the radiation-magnetohydrodynamic-thermochemical ``super-zoom'' simulations of accreting supermassive black holes from \citet{hopkins:multiphase.mag.dom.disks} (\paperone), using a time-dilation method \citep{hopkins.most:2025.time.dilation.methods.extreme.timestepping} to evolve the system for $\sim 10^{9.5}\,G M_{\rm BH}/c^{3}$ ($\sim 10^{4}\,$yr), long enough for AGN radiation and outflows to propagate from the innermost stable circular orbit ($\sim 3\,R_{g}$) to the BH radius of influence ($\sim$\,few pc). We study three fiducial accretion rates spanning $\dot{m} \sim 0.001$ to $\sim 30$ (Eddington-normalized), encompassing the transition from low-luminosity to super-Eddington accretion. Our main findings are:

\begin{enumerate}[leftmargin=*]

\item \textit{Magnetically-dominated disks are persistent:} The hyper-magnetized, flux-frozen accretion disks found in \paperone\ and earlier work \citep{hopkins:superzoom.disk,hopkins:superzoom.analytic} maintain quasi-steady-state over timescales orders of magnitude longer than previously simulated, with resolved thermal scale-heights. This argues these are not transient structures but represent a long-term quasar accretion state.

\item \textit{A state transition at low $\dot{m}$:} In our lowest-$\dot{m} \sim 0.001$ case, the inner accretion flow transitions from a cold, magnetically-dominated thin disk to a hot ($T \sim T_{\rm vir}$), geometrically-thick ($H/R \sim 0.3-0.5$) disk with $\beta \sim 1$, consistent with ADAF/RIAF models. The accretion rate scaling changes from $\dot{M} \propto R^{0.1}$ (weak outflow correction) at high $\dot{m}$ to closer to $\dot{M} \propto R^{1/2}$.

\item \textit{Emergent radiation:} The radiative efficiency $\epsilon_{r}$ peaks at $\sim 0.1$ for $0.01 \lesssim \dot{m} \lesssim 1$, and is 
consistent with standard expectations that it decreases $\propto \dot{m}$ below $\dot{m} \lesssim 0.01$ (radiatively inefficient) and $\propto \dot{m}^{-1}$ above $\dot{m} \gtrsim 1$ (photon trapping). At $\dot{m} \gtrsim 0.01$, the escaping spectrum is dominated by quasi-thermal emission at $T_{\rm rad,\,eff} \sim 10^{4}-10^{5}\,$K with infrared re-emission from the torus, consistent with observed quasar SEDs. At $\dot{m} \ll 0.01$, the inner disk becomes optically thin.

\item \textit{Radiation-pressure-driven outflows at high $\dot{m}$:} The dominant outflow mechanism at $\dot{m} \gtrsim 0.01$ is radiation pressure, with preferred launching radii at $\sim 100-1000\,R_{g}$ where opacity transitions boost the coupling. Magnetocentrifugal winds are subdominant. 
The winds emerge in a broad bicone covering $\sim 1/2$ the sky; the hottest, fastest winds are at smaller distances and larger heights (closer-to-polar). Outflow momentum-loading scales as $\sim L/c$ for $\dot{m} \lesssim 1$, increasing to $\sim 10\,L/c$ at $\dot{m} \gg 1$. Characteristic velocities of $\sim 10^{2}-10^{5}\,{\rm km\,s^{-1}}$ are consistent with observed NLR/BLR/BAL/UFO outflows.

\item \textit{Sub-grid prescriptions:} We propose simple prescriptions (\S~\ref{sec:subgrid}) for the accretion rate, radiative efficiency, and outflow momentum/energy-loading as a function of the accretion rate at the ROI, suitable for implementation in cosmological and galaxy-scale simulations. These define three qualitative feedback regimes: low-density/radio mode, intermediate/self-regulating quasar mode, and high-density/smothered hyper-Eddington mode.

\end{enumerate}

\acknowledgments{Support for PFH was provided by a Simons Investigator Grant \&\ NSF-AAG grant \#2509270. Numerical calculations were run on Frontera (LRAC AST20016) and the resources of the Texas Advanced Computing Center (TACC), including the Vista GPU system (NAIRR Startup NAIRR260139), and on the Caltech Resnick HPC cluster, under allocations AST20016 and AST21010. Generative AI (Claude Opus 4.6-4.8, Fable 5, and GPT Codex 5.5) was used in testing and debugging the code implementation in \GIZMO\ and in the writing and editing of this manuscript.}

\bibliographystyle{mn2e}
\bibliography{ms_extracted}

\end{document}